\documentclass[preprint,superscriptaddress,footinbib]{revtex4}
  \usepackage{amssymb,amsmath,graphicx,subfigure,color,rotating}
  \pdfoutput=1
  \usepackage[pdftex,colorlinks]{hyperref}
  \usepackage[left]{lineno}
  \allowdisplaybreaks[4]
  \begin{document}

  \title{Purely leptonic decays of the ground
         charged vector mesons}
  \author{Yueling Yang}
  \affiliation{Institute of Particle and Nuclear Physics,
              Henan Normal University, Xinxiang 453007, China}
  \author{Zhenglin Li}
  \affiliation{Institute of Particle and Nuclear Physics,
              Henan Normal University, Xinxiang 453007, China}
  \author{Kang Li}
  \affiliation{Institute of Particle and Nuclear Physics,
              Henan Normal University, Xinxiang 453007, China}
  \author{Jinshu Huang}
  \affiliation{School of Physics and Electronic Engineering,
              Nanyang Normal University, Nanyang 473061, China}
  \author{Junfeng Sun}
  \affiliation{Institute of Particle and Nuclear Physics,
              Henan Normal University, Xinxiang 453007, China}

  \begin{abstract}
  The study of the purely leptonic decays of the ground
  charged vector mesons is very interesting and
  significant in determining the CKM matrix elements,
  obtaining the decay constant of vector mesons,
  examining the lepton flavor universality,
  and searching for new physics beyond the standard model.
  These purely leptonic decays of the ground charged vector
  mesons are induced by the weak interactions within the
  standard model, and usually have very small branching ratios,
  ${\cal B}({\rho}^{-}{\to}{\ell}^{-}{\nu}_{\ell})$ ${\sim}$
  ${\cal O}(10^{-13})$,
  ${\cal B}(K^{{\ast}-}{\to}{\ell}^{-}{\nu}_{\ell})$ ${\sim}$
  ${\cal O}(10^{-13})$,
  ${\cal B}(D_{d}^{{\ast}-}{\to}{\ell}^{-}{\nu}_{\ell})$ ${\sim}$
  ${\cal O}(10^{-10})$,
  ${\cal B}(B_{u}^{{\ast}-}{\to}{\ell}^{-}{\nu}_{\ell})$ ${\sim}$
  ${\cal O}(10^{-10})$,
  ${\cal B}(D_{s}^{{\ast}-}{\to}{\ell}^{-}{\nu}_{\ell})$ ${\sim}$
  ${\cal O}(10^{-6})$ and
  ${\cal B}(B_{c}^{{\ast}-}{\to}{\ell}^{-}{\nu}_{\ell})$ ${\sim}$
  ${\cal O}(10^{-6})$.
  Inspired by the potential prospects of LHCb, Belle-II, STCF,
  CEPC and FCC-ee experiments, we discussed the probabilities
  of experimental investigation on these purely leptonic decays.
  It is found that the measurements of these decays might be
  possible and feasible with the improvement of data statistics,
  analytical technique, and measurement precision in the future.
  (1) With the hadron-hadron collisions, the purely leptonic decays
  of ${\rho}^{-}$, $K^{{\ast}-}$, $D_{d,s}^{{\ast}-}$ and
  $B_{u,c}^{{\ast}-}$ mesons might be accessible at LHC experiments.
  (2) With the $e^{+}e^{-}$ collisions, the purely leptonic decays
  of $D_{d,s}^{{\ast}-}$ and $B_{u,c}^{{\ast}-}$ mesons
  might be measurable with over $10^{12}$ $Z^{0}$ bosons
  available at CEPC and FCC-ee experiments.
  In addition, the $D_{d,s}^{{\ast}-}$ ${\to}$
  ${\ell}^{-}{\nu}_{\ell}$ decays could also be studied
  at Belle-II and SCTF experiments.

  {\color{blue} Published :}
   \href{https://doi.org/10.1140/epjc/s10052-021-09908-w}
       {Eur. Phys. J. C 81, 1110 (2021).}
  \end{abstract}

  \maketitle

  \section{Introduction}
  \label{sec01}
   In the quark model \cite{pdg2020,pl.8.214,zweig},
   mesons are generally regarded as bound states of the valence
   quark $q$ and antiquark $\bar{q}^{\prime}$.
   The classifications of mesons are usually based on the spin-parity
   quantum number $J^{P}$ of the $q\bar{q}^{\prime}$ system.
   The spin $J$ of meson is given by the relation ${\vert}L-S{\vert}$
   ${\le}$ $J$ ${\le}$ ${\vert}L+S{\vert}$.
   The orbital angular momentum and total spin of the
   $q\bar{q}^{\prime}$ system are respectively $L$ and $S$,
   where $S$ $=$ $0$ for antiparallel quark spins, and $S$ $=$ $1$
   for parallel quark spins.
   By convention, quarks have a positive parity and antiquarks
   have a negative parity.
   Hence, the parity of meson is $P$ $=$ $(-1)^{L+1}$.
   The $L$ $=$ $0$ states are the ground-state pseudoscalars
   with $J^{P}$ $=$ $0^{-}$ and vectors with $J^{P}$ $=$ $1^{-}$.
   Both quarks and leptons are fermions with spin $S$ $=$ $1/2$.
   Mesons are composed of a pair of fermions --- quark and antiquark,
   therefore, they could in principle decay into a pair of fermions,
   for example, lepton and antilepton.
   The experimental observation of the two-body purely leptonic
   decays of mesons could be a clear and characteristic manifestation
   of the quark model. These leptonic decays provide us with
   valuable opportunities to fully investigate the microstructure
   and properties of mesons.
   The study of two-body purely leptonic decays of mesons is very
   interesting and significant.

   The valence quarks of the electrically charged mesons must
   have different flavors.
   Within the standard model (SM) of elementary particles,
   the purely leptonic decays of the charged mesons (PLDCM)
   are typically induced by the tree-level exchange of the
   gauge bosons $W$, the quanta of the weak interaction fields.
   Up to today, the masses of all the experimentally observed
   mesons are much less than those of $W$ bosons.
   Consequently, the massive $W$ bosons are virtual propagators
   rather than physical particles in the true picture of PLDCM.
   Phenomenologically, by integrating out the contributions
   from heavy dynamical degrees of freedom such as the $W$ fields,
   PLDCM can be properly described by the low-energy effective
   theory in analogy with the Fermi theory for ${\beta}$ decays.
   Considering the fact that leptons are free from the strong
   interactions, the corresponding effective Hamiltonian
   \cite{RevModPhys.68.1125} for PLDCM could be written as
   the product of quark currents and leptonic currents,
   \begin{equation}
  {\cal H}_{\rm eff}\, =\,
   \frac{G_{F}}{\sqrt{2}}\, V_{q_{1}q_{2}}\,
   \big[ \bar{q}_{1}{\gamma}_{\mu}(1-{\gamma}_{5})q_{2} \big]\,
   \big[ \bar{\ell}\,{\gamma}^{\mu}(1-{\gamma}_{5}){\nu}_{\ell} \big]
      + {\rm h.c.}
   \label{eq:hamilton},
   \end{equation}
   where the contributions of the $W$ bosons are embodied in the Fermi
   coupling constant $G_{F}$ ${\simeq}$
   $1.166{\times}10^{-5}\,{\rm GeV}^{-2}$ \cite{pdg2020},
   and $V_{q_{1}q_{2}}$ is the Cabibbo-Kobayashi-Maskawa
   (CKM) \cite{Cabibbo,Kobayashi} matrix element between the
   quarks in the charged mesons.
   The decay amplitudes can be written as,
   \begin{equation}
  {\langle}{\ell}\bar{\nu}_{\ell}{\vert}{\cal H}_{\rm eff}
  {\vert}M{\rangle}\, =\,
   \frac{G_{F}}{\sqrt{2}}\, V_{q_{1}q_{2}}\,
  {\langle}{\ell}\bar{\nu}_{\ell}{\vert}
   \bar{\ell}\,{\gamma}^{\mu}(1-{\gamma}_{5}){\nu}_{\ell}
  {\vert}0{\rangle}\, {\langle}0{\vert}
   \bar{q}_{1}{\gamma}_{\mu}(1-{\gamma}_{5})q_{2}
  {\vert}M{\rangle}
   \label{eq:amplitude}.
   \end{equation}
   The leptonic part of amplitudes can be calculated reliably
   with the perturbative theory.
   The hadronic matrix elements (HMEs) interpolating the diquark
   currents between the mesons concerned and the vacuum can be
   parameterized by the decay constants.

   With the conventions of Refs. \cite{JHEP.0605.004,JHEP.0703.069},
   the HMEs of diquark currents are defined as,
   \begin{eqnarray}
  {\langle}0{\vert} \bar{q}_{1}(0)\,{\gamma}_{\mu}\,q_{2}(0)
  {\vert}P(k){\rangle} &=& 0
   \label{eq:hme-pseudoscalar-v},
   \\
  {\langle}0{\vert} \bar{q}_{1}(0)\,{\gamma}_{\mu}\,
  {\gamma}_{5}\,q_{2}(0){\vert}P(k){\rangle} &=& i\,f_{P}\,k_{\mu}
   \label{eq:hme-pseudoscalar-a},
   \\
  {\langle}0{\vert} \bar{q}_{1}(0)\,{\gamma}_{\mu}\,q_{2}(0)
  {\vert}V(k,{\epsilon}){\rangle} &=& f_{V}\,m_{V}\,{\epsilon}_{\mu}
   \label{eq:hme-vector-v},
   \\
  {\langle}0{\vert} \bar{q}_{1}(0)\,{\gamma}_{\mu}\,
  {\gamma}_{5}\,q_{2}(0){\vert}V(k,{\epsilon}){\rangle} &=& 0
   \label{eq:hme-vector-a},
   \end{eqnarray}
   where the nonperturbative parameters of $f_{P}$ and $f_{V}$
   are the decay constants of pseudoscalar $P$ and vector $V$
   mesons, respectively; and $m_{V}$ and ${\epsilon}_{\mu}$ are
   the mass and polarization vector, respectively.
   To the lowest order, the decay widths are written as,
   \begin{eqnarray}
  {\Gamma}(P{\to}{\ell}\,\bar{\nu}_{\ell}) &=&
   \frac{G_{F}^{2}}{8{\pi}}\,{\vert}V_{q_{1}q_{2}}{\vert}^{2}\,
   f_{P}^{2}\,m_{P}\, m_{\ell}^{2}\, \Big(
   1-\frac{m_{\ell}^{2}}{m_{P}^{2}} \Big)^{2}
   \label{eq:decay-width-pseudoscalar},
   \\
  {\Gamma}(V{\to}{\ell}\,\bar{\nu}_{\ell}) &=&
   \frac{G_{F}^{2}}{12{\pi}}\,{\vert}V_{q_{1}q_{2}}{\vert}^{2}\,
   f_{V}^{2}\,m_{V}^{3}\, \Big( 1-\frac{m_{\ell}^{2}}{m_{V}^{2}}
   \Big)^{2}\, \Big( 1+\frac{m_{\ell}^{2}}{2\,m_{V}^{2}} \Big)
   \label{eq:decay-width--vector},
   \end{eqnarray}
   where $m_{P}$ and $m_{\ell}$ are the masses of the charged
   pseudoscalar meson and lepton, respectively.

   It is clearly seen from the above formula that the highly precise
   measurements of PLDCM will allow the relatively accurate
   determinations of the product of the decay constants and
   CKM elements, ${\vert}V_{q_{1}q_{2}}{\vert}f_{P,V}$.
   Theoretically, the decay constants are nonperturbative parameters,
   and they are closely related with the $\bar{q}_{1}q_{2}$
   wave functions at the origin which cannot be computed from
   first principles.
   There still exist some discrepancies among theoretical
   results of the decay constants with different methods, such
   as the potential model, QCD sum rules, lattice QCD, and so on.
   If the magnitudes of CKM element ${\vert}V_{q_{1}q_{2}}{\vert}$
   are fixed to the values of Ref. \cite{pdg2020},
   the decay constants $f_{P,V}$ will be experimentally measured,
   and be used to seriously examine the different calculations
   on the decay constants with various theoretical models.
   Likewise, if the decay constants $f_{P,V}$ are well known to
   sufficient precision, the magnitudes of the corresponding
   CKM element will be experimentally determined, and provide
   complementary information to those from other processes.
   Within SM, the $P$ ${\to}$ ${\ell}\bar{\nu}_{\ell}$ and
   $V$ ${\to}$ ${\ell}\bar{\nu}_{\ell}$ decays are induecd
   by the axial-vector current of Eq.(\ref{eq:hme-pseudoscalar-a})
   and vector current of Eq.(\ref{eq:hme-vector-v}),
   respectively; and the electroweak interactions assign the
   vector-minus-axial-vector ($V-A$) currents to the $W$ bosons.
   The CKM elements determined from two different and complementary
   parts of the electroweak interactions, charged vector and
   axial-vector currents, could be independently examined.
   The latest CKM elements determined by PLDCM, such as
   ${\vert}V_{us}{\vert}$, ${\vert}V_{cd}{\vert}$ and
   ${\vert}V_{cs}{\vert}$, differ somewhat from those by
   exclusive and inclusive semileptonic meson decays \cite{pdg2020}.
   The CKM elements extracted from various processes can be
   combined to test the electroweak characteristic
   charged-current $V-A$ interactions.

   Within SM, the lepton-gauge-boson electroweak gauge couplings
   are generally believed to be universal and process independent,
   which is called lepton flavor universality (LFU).
   However, there are some hints of LFU discrepancies between
   SM predictions and experimental measurements, such as the
   ratios of branching fractions of semileptonic $B$ decays
   $R(D^{(\ast)})$ ${\equiv}$
   ${\cal B}(\bar{B}{\to}D^{(\ast)}{\tau}\bar{\nu}_{\tau})/
    {\cal B}(\bar{B}{\to}D^{(\ast)}{\ell}\bar{\nu}_{\ell})$
   with ${\ell}$ $=$ $e/{\mu}$ \cite{pdg2020}.
   The LFU validity can be carefully investigated through
   the PLDCM processes.
   Beyond SM, some possible new heavy particles accompanied
   with novel interactions, such as the charged higgs bosons, 
   would affect PLDCM and LFU, and might
   lead to detectable effects.
   So PLDCM provide good arenas to search for the smoking
   gun of new physics (NP) beyond SM.

   By considering the angular momentum conservation and the
   final states including a left-handed neutrino or
   right-handed antineutrino, the purely leptonic decay
   width of charged pseudoscalar meson,
   Eq.(\ref{eq:decay-width-pseudoscalar}),
   is proportional to the square of the lepton mass.
   This is called helicity suppression.
   While there is no helicity suppression for the purely
   leptonic decay of charged vector meson (PLDCV).
   From the analytical expressions of
   Eq.(\ref{eq:decay-width-pseudoscalar}) and
   Eq.(\ref{eq:decay-width--vector}),
   the decay width of pseudoscalar meson is suppressed
   by the factor $m_{\ell}^{2}/m_{P}^{2}$ compared with
   that of vector meson.
   What's more, both the masses and the decay constants of
   vector mesons are relatively larger than those of corresponding
   pseudoscalar mesons, which would result in an enhancement
   of the decay widths for vector mesons.
   Of course, the vector mesons decay dominantly through
   the strong and/or electromagnetic interactions.
   The branching ratios for the PLDCV weak decays are usually very small,
   sometimes even close to the accessible limits of the existing
   and the coming experiments.

   Inspired by the potential prospects of the future high-intensity
   and high-energy frontiers, along with the noticeable increase of
   experimental data statistics, the remarkable improvement of
   analytical technique 
   and the continuous enhancement of measurement precision,
   the carefully experimental study of PLDCV might be possible
   and feasible.
   In this paper, we will focus on the PLDCV within SM
   to just provide a ready reference.
   The review of the purely leptonic decays of charged
   pseudoscalar mesons can be found in Ref. \cite{pdg2020}.

  \section{${\rho}^{-}$ ${\to}$ ${\ell}^{-}\bar{\nu}_{\ell}$ decays}
  \label{sec-rho}
  The mass of the ${\rho}^{\pm}$ meson, $m_{\rho}$ $=$ $775.11(34)$ MeV
  \cite{pdg2020}, is much larger than that of two-pion pair.
  The rate of the ${\rho}$ meson decay into two pions via the
  strong interactions is almost 100\%, which results in the very
  short lifetime ${\tau}_{\rho}$ ${\sim}$ $4.4{\times}10^{-24}$ s
  \cite{pdg2020}.
  The direct measurements of the electroweak properties of the
  ${\rho}$ meson would definitely be very challenging.
  It is evident from Eq.(\ref{eq:decay-width--vector})
  that the parameter of ${\vert}V_{ud}{\vert}\,f_{\rho}$ could be
  experimentally determined from the observations of decay widths
  for the ${\rho}^{-}$ ${\to}$ ${\ell}^{-}\bar{\nu}_{\ell}$ decays
  (if it is not specified, the corresponding charge-conjugation
  processes are included in this paper),
  with the coupling constant $G_{F}$, the masses of
  lepton $m_{\ell}$ and meson $m_{\rho}$.

\begin{figure}[ht]
  \includegraphics[width=0.55\textwidth]{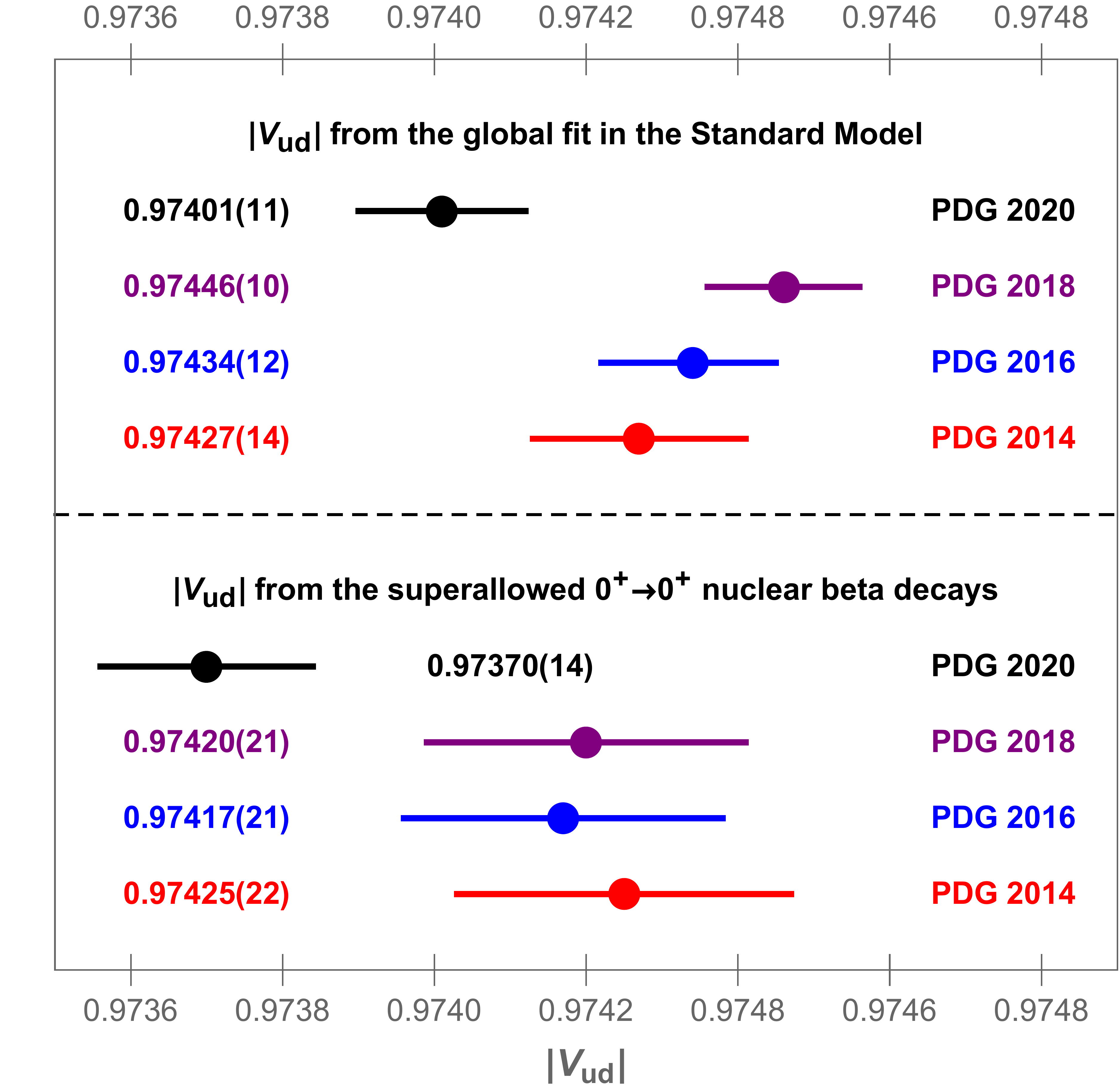}
  \caption{The values of the CKM element ${\vert}V_{ud}{\vert}$
  from Particle Data Group (PDG).}
  \label{fig:vud}
  \end{figure}

  The precise values of the CKM element ${\vert}V_{ud}{\vert}$ in
  ascending order of measurement accuracy mainly come from ${\beta}$
  transitions between the super-allowed nuclear analog states with
  quantum number of both $J^{P}$ $=$ $0^{+}$ and isospin $I$ $=$ $1$,
  between mirror nuclei with $I$ $=$ $1/2$,
  between neutron and proton,
  between charged and neutral pions \cite{CKM2016-vud}.
  These four results for 
  ${\vert}V_{ud}{\vert}$ are basically
  consistent with one another.
  The result of the super-allowed $0^{+}$ ${\to}$ $0^{+}$ nuclear ${\beta}$
  transitions has an uncertainty a factor of about 10
  smaller than the other results, and thus dominates the weighted
  average value \cite{CKM2016-vud}.
  The best value from super-allowed nuclear ${\beta}$ transitions is
  ${\vert}V_{ud}{\vert}$ $=$ $0.97370(14)$ \cite{pdg2020}, which is
  smaller compared with the 2018 value ${\vert}V_{ud}{\vert}$ $=$
  $0.97420(21)$ \cite{pdg2018}, as illustrated Fig. \ref{fig:vud}.
  This reduction of the value of ${\vert}V_{ud}{\vert}$
  leads to a slight deviation from the first row unitarity
  requirement ${\vert}V_{ud}{\vert}^{2}$ $+$ ${\vert}V_{us}{\vert}^{2}$
  $+$ ${\vert}V_{ub}{\vert}^{2}$ $=$ $1$.
  The current precision of the CKM element ${\vert}V_{ud}{\vert}$
  is about 0.01\%. The latest value from the global fit in SM,
  ${\vert}V_{ud}{\vert}$ $=$ $0.97401(11)$ \cite{pdg2020},
  will be used in our calculation.

   \begin{table}[ht]
   \caption{Decay constant $f_{\rho}$ (in the unit of MeV)
   obtained from the diffractive vector meson production
   using deep inelastic scattering (DVP), QCD sum rules (SR),
   relativistic quark model (RQM), light-front quark model (LFQM),
   lattice QCD (LQCD) and so on.}
   \label{tab:rho-decay-constant}
   \begin{ruledtabular}
   \begin{tabular}{lccccc}
   DVP & $143$ \cite{PhysRevD.69.094013} \footnotemark[1]
       & $147$ \cite{PhysRevD.69.094013} \footnotemark[2]
       & $153$ \cite{PhysRevD.69.094013} \footnotemark[3]
       & $161$ \cite{PhysRevD.69.094013} \footnotemark[4]
       & $211$ \cite{PhysRevD.94.074018} \\
   SR  & $201{\pm}5$ \cite{plb.436.351}\footnotemark[5]
       & $205{\pm}10$ \cite{plb.436.351} \footnotemark[6]
       & $194$ \cite{plb.436.351} \footnotemark[7]
       & $198{\pm}7$ \cite{PhysRevD.58.094016}
       & $206{\pm}7$ \cite{JHEP.0604.046} \\
  RQM  & $168.3$ \cite{epja.24.411} \footnotemark[8]
       & $151.3$ \cite{epja.24.411} \footnotemark[9]
       & $190.1$ \cite{epja.24.411} \footnotemark[10]
       & $175.4$ \cite{epja.24.411} \footnotemark[11]
       & $219$ \cite{plb.635.93} \\
  LFQM & $246$ \cite{PhysRevD.75.034019} \footnotemark[12]
       & $215$ \cite{PhysRevD.75.034019}  \footnotemark[13]
       & $205$ \cite{PhysRevC.92.055203} \footnotemark[12]
       & $166^{+2}_{-4}$ \cite{cpc.42.073102} \footnotemark[14]
       & $210{\pm}6$ \cite{cpc.42.073102} \footnotemark[15] \\
  LFQM & $215{\pm}5$ \cite{jpg.39.025005}
       & $211{\pm}1$ \cite{PhysRevD.98.114018}
       & $242^{+23}_{-24}$ \cite{PhysRevD.100.014026}  \\
  LQCD  & $239.4{\pm}7.3$ \cite{PhysRevD.65.054505}
        & $210{\pm}15$ \cite{PTP.119.599}
        & $239{\pm}18$ \cite{PhysRevD.80.054510}
        & $199{\pm}4$ \cite{JHEP.1704.082}
        & $208.5{\pm}5.6$ \cite{cpc.42.063102} \\
  other & $490$ \cite{plb.635.93}
        & $207$ \cite{PhysRevC.60.055214}
        & $254$ \cite{pan.79.444}
         \\
  \end{tabular}
  \end{ruledtabular}
  \footnotetext[1]{With light-cone wave functions and parameters of Ref. \cite{epjc.22.655}.}
  \footnotetext[2]{With light-cone wave functions and parameters of Ref. \cite{PhysRevD.60.114023}.}
  \footnotetext[3]{With Gaussian wave functions and parameters of Ref. \cite{epjc.22.655}.}
  \footnotetext[4]{With Gaussian wave functions and parameters of Ref. \cite{PhysRevD.60.114023}.}
  \footnotetext[5]{With nonlocal condensates like functions.}
  \footnotetext[6]{With Ball-Braun wave functions \cite{PhysRevD.54.2182}.}
  \footnotetext[7]{With Chernyak-Zhitnitsky wave functions \cite{PhysRept.112.173}.}
  \footnotetext[8]{With Gaussian spatial wave functions and adjusted parameters of Ref. \cite{epja.24.411}.}
  \footnotetext[9]{With Gaussian spatial wave functions and parameters of Ref. \cite{plb.602.212}.}
  \footnotetext[10]{With rational spatial wave functions and adjusted parameters of Ref. \cite{epja.24.411}.}
  \footnotetext[11]{With rational spatial wave functions and parameters of Ref. \cite{plb.602.212}.}
  \footnotetext[12]{With Coulomb plus linear potential model.}
  \footnotetext[13]{With Coulomb plus harmonic oscillator potential model.}
  \footnotetext[14]{With a dilation parameter ${\kappa}$ $=$ $0.54$ GeV.}
  \footnotetext[15]{With a dilation parameter ${\kappa}$ $=$ $0.68$ GeV.}
  \end{table}

  The decay constant $f_{\rho}$ is an very important
  characteristics of the ${\rho}$ meson.
  Compared with the CKM element ${\vert}V_{ud}{\vert}$,
  the present precision of decay constant $f_{\rho}$
  is still not very high and needs to be improved.
  Theoretically, the estimations from different methods are
  more or less different from each other and even calculations
  with the same method sometimes give the diverse results.
  Some theoretical estimations on the decay constant
  $f_{\rho}$ are presented in Table \ref{tab:rho-decay-constant}.
  Experimentally, the decay constant $f_{\rho}$ can be
  obtained from the 1-prong hadronic ${\tau}^{\pm}$ ${\to}$
  ${\rho}^{\pm}{\nu}_{\tau}$ decay.
  The partial width for the ${\tau}$ ${\to}$ $V{\nu}_{\tau}$
  decay is given by Ref. \cite{JHEP.1504.101},
   \begin{equation}
  {\Gamma}({\tau}{\to}V{\nu})\, =\,
  {\cal S}\,\frac{G_{F}^{2}\,m_{\tau}^{3}}{16\,{\pi}}\,
  {\vert}V_{q_{1}q_{2}}{\vert}^{2}\, f_{V}^{2}\,
   \Big( 1-\frac{m_{V}^{2}}{m_{\tau}^{2}} \Big)^{2}
   \Big( 1+\frac{2\,m_{V}^{2}}{m_{\tau}^{2}} \Big),
   \label{eq:tau-decay-width}
   \end{equation}
  where the factor ${\cal S}$ $=$ $1.0154$ includes the electroweak
  corrections \cite{JHEP.1504.101,PhysRevLett.61.1815,PhysRevD.42.3888}.
  With the mass $m_{\tau}$ $=$ $1776.86(12)$ MeV and
  lifetime ${\tau}_{\tau}$ $=$ $290.3(5)$ fs \cite{pdg2020},
  and branching ratio ${\cal B}({\tau}{\to}{\rho}{\nu})$
  $=$ $25.19(33)$ \% \cite{pdg2020}, one can easily extract the
  decay constant $f_{\rho}^{\rm exp}$ $=$ $207.7{\pm}1.6$ MeV,
  which agrees well with the latest numerical simulation result
  from lattice QCD $f_{\rho}$ $=$ $208.5{\pm}5.5{\pm}0.9$ MeV
  \cite{cpc.42.063102}.
  The more accurate 
  decay constant $f_{\rho}^{\rm exp}$
  will be used in our calculation.

  For the ${\rho}^{-}$ ${\to}$ ${\ell}^{-}\bar{\nu}_{\ell}$ decays,
  one can obtain the PLDCV partial decay widths with
  Eq.(\ref{eq:decay-width--vector}) and the corresponding
  branching ratios with the full width ${\Gamma}_{\rho}$
  $=$ $149.1{\pm}0.8$ MeV \cite{pdg2020},
   \begin{eqnarray}
  {\Gamma}({\rho}^{-}{\to}e^{-}\bar{\nu}_{e})
   &=& 68.8{\pm}1.2\, {\mu}\,{\rm eV}
   \label{eq:width-rho-e}, \\
  {\Gamma}({\rho}^{-}{\to}{\mu}^{-}\bar{\nu}_{\mu})
   &=& 66.9{\pm}1.1\, {\mu}\,{\rm eV}
   \label{eq:width-rho-mu}, \\
   {\cal B}({\rho}^{-}{\to}e^{-}\bar{\nu}_{e})
   &=& (4.6{\pm}0.1){\times}10^{-13}
   \label{eq:br-rho-e}, \\
  {\cal B}({\rho}^{-}{\to}{\mu}^{-}\bar{\nu}_{\mu})
   &=& (4.5{\pm}0.1){\times}10^{-13}
   \label{eq:br-rho-mu},
   \end{eqnarray}
  where the uncertainties come from the uncertainties of
  mass $m_{\rho}$, decay constant $f_{\rho}$ and CKM element
  ${\vert}V_{ud}{\vert}$, and additional decay width
  ${\Gamma}_{\rho}$ for branching ratios.
  Clearly, the branching ratios are very small.
  Given the identification efficiency and pollution from %
  background, the ${\rho}^{-}$ ${\to}$ ${\ell}^{-}\bar{\nu}_{\ell}$
  decays might be measured only with more than $10^{14}$
  ${\rho}^{\pm}$ events available.

  There are at least three possible ways to experimentally produce
  the charged ${\rho}$ mesons in the electron-position collisions,
  (a) the prompt pair production $e^{+}e^{-}$ ${\to}$
  ${\rho}^{+}{\rho}^{-}$,
  (b) the pair production via $V$ decay $1^{--}$ ${\to}$
  ${\rho}^{+}{\rho}^{-}$,
  and (c) the single production via $V$ decay $1^{--}$ ${\to}$
  ${\rho}^{\pm}h^{\mp}$.
  The cross section ${\sigma}(e^{+}e^{-}{\to}{\rho}^{+}{\rho}^{-})$
  has been determined by the BaBar group to be
  $19.5{\pm}1.6{\pm}3.2$ fb near the center-of-mass
  energy $\sqrt{s}$ $=$ $10.58$ GeV \cite{PhysRevD.78.071103}.
  Assuming the production cross section ${\sigma}$ ${\propto}$
  $1/s$ \cite{PhysRevC.75.065202,plb.763.87},
  it could be speculated that
  ${\sigma}(e^{+}e^{-}{\to}{\rho}^{+}{\rho}^{-})$
  ${\sim}$ $230$ fb near $\sqrt{s}$ $=$ $3.1$ GeV.
  There would be only about $10^{6}$ ${\rho}^{+}{\rho}^{-}$ pairs
  with a data sample of $50$ ${\rm ab}^{-1}$ \cite{PTEP.2019.123C01}
  near $\sqrt{s}$ ${\approx}$ $m_{\Upsilon(4S)}$ at the Belle-II detector
  or a data sample of $10$ ${\rm ab}^{-1}$ \cite{epjconf.212.01010}
  near $\sqrt{s}$ ${\approx}$ $m_{J/{\psi}}$ with the future super-tau-charm
  factory like  STCF or SCTF \cite{PhysRevLett.127.012003,STCF,SCTF}.
  The charge ${\rho}$ mesons can in principle be
  produced from the ${\Upsilon}(4S)$, $J/{\psi}$ and ${\phi}$ decays.
  The branching ratios are
   \begin{eqnarray}
  {\cal B}({\Upsilon}(4S){\to}{\rho}^{+}{\rho}^{-})
   &<& 5.7{\times}10^{-6}\
   \text{\cite{PhysRevD.78.071103}}
   \label{eq:br-y4s-rho-pair}, \\
  {\cal B}(J/{\psi}{\to}{\rho}^{+}{\rho}^{-})
   &{\sim}& 10^{-3}\
   \text{\cite{PhysRevD.32.2961,pdg2020}}
   \label{eq:br-j/psi-rho-pair}, \\
  {\cal B}(J/{\psi}{\to}{\rho}^{\pm}{\pi}^{\mp})
   &{\sim}& 10^{-2}\
   \text{\cite{pdg2020}}
   \label{eq:br-j/psi-rho-pi}, \\
  {\cal B}({\phi}{\to}{\rho}^{\pm}{\pi}^{\mp})
   &{\sim}& 10^{-1}\
   \text{\cite{pdg2020}}
   \label{eq:br-phi-rho-pi},
   \end{eqnarray}
  where the brancing ratio ${\cal B}(J/{\psi}{\to}{\rho}^{+}{\rho}^{-})$
  is assumed to be the same order of magnitude as
  ${\cal B}(J/{\psi}{\to}K^{{\ast}+}K^{{\ast}-})$
  ${\sim}$ $10^{-3}$ \cite{pdg2020} from the phenomenological
  analysis based on the flavor-$SU(3)$ symmetry \cite{PhysRevD.32.2961}.
  Now, there are $7.7{\times}10^{8}$ ${\Upsilon}(4S)$ events at Belle
  \cite{epjc74.3026}, $10^{10}$ $J/{\psi}$ events \cite{PhysRevLett.127.012003}
  at BES-III, and $2.4{\times}10^{10}$ ${\phi}$ events at KLOE/KLOE-2
  \cite{KLOE2.2020} available. 
  It is expected that only about $5{\times}10^{10}$ ${\Upsilon}(4S)$
  events \cite{PTEP.2019.123C01} and $10^{13}$ 
  $J/{\psi}$ events at SCTF or STCF \cite{PhysRevLett.127.012003}
  could be accumulated.
  It is clearly seen that unless a very significant enhancement to
  branching ratios from some NP, the experimental data on the
  ${\rho}^{\pm}$ meson are too scarce to search for the
  ${\rho}^{-}$ ${\to}$ ${\ell}^{-}\bar{\nu}_{\ell}$
  decays at the electron-position collisions in the near future,
  which result in the natural difficulties to understand 
  the ${\rho}$ meson.

  The production cross sections of prompt $J/{\psi}$ and
  $J/{\psi}$-from-b mesons in proton–proton collisions at $\sqrt{s}$
  $=$ $13$ TeV are measured by LHCb to be $15.0{\pm}0.6{\pm}0.7$
  ${\mu}$b and $2.25{\pm}0.09{\pm}0.10$ ${\mu}$b, respectively,
  \cite{JHEP.2015.10.172}.
  It is expected that some $10^{12}$ $J/{\psi}$ events could be
  accumulated at $\sqrt{s}$ $=$ $13$ TeV with an integrated
  luminosity of $300$ ${\rm fb}^{-1}$ at LHCb \cite{1808.08865}.
  There are only about $10^{10}$ ${\rho}^{\pm}$ mesons from $J/{\psi}$
  decays available for prying into the ${\rho}^{\pm}$ PLDCV decays.
  At the same time, the inclusive cross-sections for
  prompt charm production at LHCb at $\sqrt{s}$ $=$ $13$ TeV are
  measured to be ${\cal O}(1\,{\rm mb})$ \cite{JHEP.2016.03.159}.
  Analogically assuming the inclusive cross section of prompt
  ${\rho}^{\pm}$ meson production at LHCb at $\sqrt{s}$ $=$ $13$
  TeV is ${\cal O}(10\,{\rm mb})$, some $3{\times}10^{15}$
  ${\rho}^{\pm}$ events would be accumulated with an integrated
  luminosity of $300$ ${\rm fb}^{-1}$ at LHCb \cite{1808.08865}.
  Optimistically assuming the reconstruction efficiency is about
  $10\,\%$, there would be about ${\cal O}(10^{2})$ events of
  the ${\rho}^{-}$ ${\to}$ ${\ell}^{-}\bar{\nu}_{\ell}$ decays
  at LHCb, and more events with the enhanced branching ratios
  from NP contributions.
  Even through it will be very challenging for experimental
  analysis due to the complex background in hadron-hadron collisions,
  there is still a strong presumption that the ${\rho}^{-}$ ${\to}$
  ${\ell}^{-}\bar{\nu}_{\ell}$ decays could be explored and studied
  at LHC in the future.
  In addition, it is expected that an integrated luminosity exceeding
  $10$ ${\rm ab}^{-1}$ would be reached at the future HE-LHC
  experiments \cite{epjst.228.1109}. More experimental data at HE-LHC
  would make the study of the ${\rho}^{-}$ ${\to}$ ${\ell}^{-}\bar{\nu}_{\ell}$
  decays indeed feasible in hadron-hadron collisions.

  \section{$K^{{\ast}-}$ ${\to}$ ${\ell}^{-}\bar{\nu}_{\ell}$ decays}
  \label{sec-kstar}
  The parameter product ${\vert}V_{us}{\vert}\,f_{K^{\ast}}$ could
  be experimentally determined from the $K^{{\ast}-}$ ${\to}$
  ${\ell}^{-}\bar{\nu}_{\ell}$ decays using
  Eq.(\ref{eq:decay-width--vector}).
  Like the ${\rho}^{\pm}$ meson, the mass of the $K^{{\ast}{\pm}}$
  meson, $m_{K^{{\ast}{\pm}}}$ $=$ $895.5(8)$ MeV, is above the
  threshold of $K{\pi}$ pair, and the partial branching ratio of
  the $K^{\ast}$ meson decay into $K{\pi}$ pair via the strong
  interactions is almost 100\% \cite{pdg2020}.
  It is not hard to imagine that the very short lifetime ${\tau}_{K^{\ast}}$
  ${\sim}$ $1.4{\times}10^{-23}$ s would enable the measurements of the
  electroweak properties of the $K^{\ast}$ meson to be
  very challenging or nearly impossible.

\begin{figure}[ht]
  \includegraphics[width=0.55\textwidth]{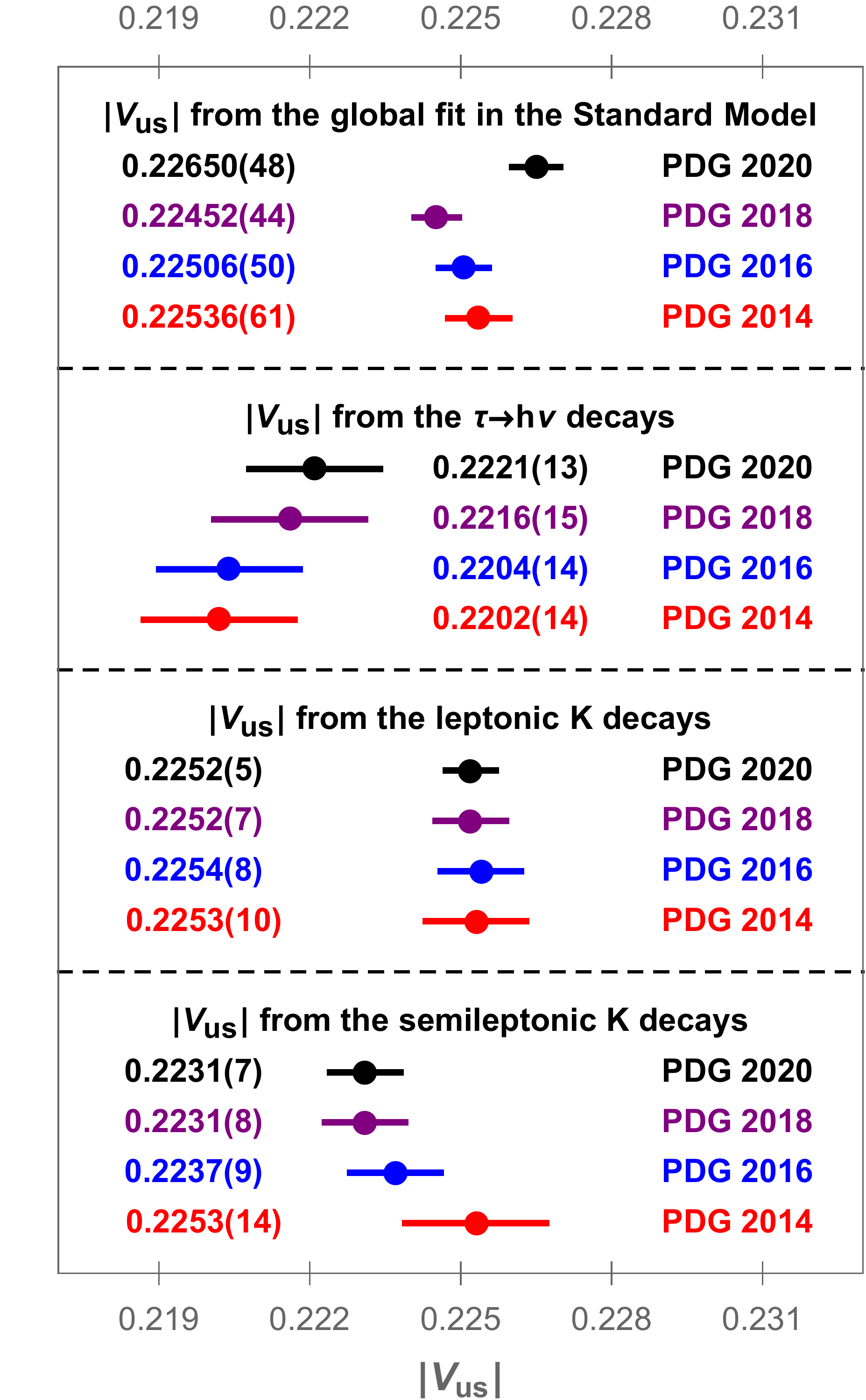}
  \caption{The values of the CKM element ${\vert}V_{us}{\vert}$
  from PDG.}
  \label{fig:vus}
  \end{figure}

  The CKM element ${\vert}V_{us}{\vert}$ ${\simeq}$ ${\lambda}$ up
  to the order of ${\cal O}({\lambda}^{6})$, where ${\lambda}$ is
  a Wolfenstein parameter.
  The current precision of the CKM element ${\vert}V_{us}{\vert}$
  from purely leptonic and semileptonic $K$ meson decays and
  hadronic ${\tau}$ decays are 0.2\%, 0.3\% and 0.6\%, respectively.
  It is seen from Fig.~\ref{fig:vus} that these three results for
  ${\vert}V_{us}{\vert}$ are not very consistent with one another.
  So if the $K^{{\ast}-}$ ${\to}$ ${\ell}^{-}\bar{\nu}_{\ell}$ decays
  could be measured, they would provide another determination and
  constraint to ${\vert}V_{us}{\vert}$.
  Probably due to the reduction of the value of ${\vert}V_{ud}{\vert}$,
  the latest value from the global fit in SM,
  ${\vert}V_{us}{\vert}$ $=$ $0.22650(48)$ \cite{pdg2020},
  is slightly larger than the 2018 value, to satisfy the
  first row unitarity requirement.

   \begin{table}[ht]
   \caption{The theoretical values of decay constant $f_{K^{\ast}}$
   (in the unit of MeV), where the legends are the same as those in
   Table \ref{tab:rho-decay-constant}.}
   \label{tab:kstar-decay-constant}
   \begin{ruledtabular}
   \begin{tabular}{lccccc}
  RQM  & $170.5$ \cite{epja.24.411} \footnotemark[1]
       & $153.1$ \cite{epja.24.411} \footnotemark[2]
       & $192.6$ \cite{epja.24.411} \footnotemark[3]
       & $177.6$ \cite{epja.24.411} \footnotemark[4]
       & $236$ \cite{plb.635.93} \\
  LFQM & $256$ \cite{PhysRevD.75.034019} \footnotemark[5]
       & $223$ \cite{PhysRevD.75.034019}  \footnotemark[6]
       & $224$ \cite{PhysRevC.92.055203} \footnotemark[5]
       & $186^{+2}_{-3}$ \cite{cpc.42.073102} \footnotemark[7]
       & $204^{+7}_{-9}$ \cite{cpc.42.073102} \footnotemark[8] \\
  LFQM & $217{\pm}5$ \cite{jpg.39.025005}
       & $223{\pm}1$ \cite{PhysRevD.98.114018}
       & $253{\pm}25$ \cite{PhysRevD.100.014026} \\
  LQCD  & $255.5{\pm}6.5$ \cite{PhysRevD.65.054505} \footnotemark[9]
        & $257.7{\pm}6.4$ \cite{PhysRevD.65.054505} \footnotemark[10]
        & $240{\pm}18$ \cite{PhysRevD.84.014505} \\
  other & $226{\pm}28$ \cite{PhysRevD.58.094016}
        & $508$ \cite{plb.635.93}
        & $241$ \cite{PhysRevC.60.055214} \\
  \end{tabular}
  \end{ruledtabular}
  \footnotetext[1]{With Gaussian spatial wave functions and adjusted parameters of Ref. \cite{epja.24.411}.}
  \footnotetext[2]{With Gaussian spatial wave functions and parameters of Ref. \cite{plb.602.212}.}
  \footnotetext[3]{With rational spatial wave functions and adjusted parameters of Ref. \cite{epja.24.411}.}
  \footnotetext[4]{With rational spatial wave functions and parameters of Ref. \cite{plb.602.212}.}
  \footnotetext[5]{With Coulomb plus linear potential model.}
  \footnotetext[6]{With Coulomb plus harmonic oscillator potential model.}
  \footnotetext[7]{With a dilation parameter ${\kappa}$ $=$ $0.54$ GeV.}
  \footnotetext[8]{With a dilation parameter ${\kappa}$ $=$ $0.68$ GeV.}
  \footnotetext[9]{The $K$ meson mass is used as input.}
  \footnotetext[10]{The ${\phi}$ meson mass is used as input.}
  \end{table}

  Some theoretical results on the decay constant $f_{K^{\ast}}$
  are presented in Table \ref{tab:kstar-decay-constant}.
  Like the case of the decay constant $f_{\rho}$,
  the model dependence of theoretical estimations on the decay
  constant $f_{K^{\ast}}$ is also obvious.
  Experimentally, the decay constant $f_{K^{\ast}}$ can be
  obtained from the hadronic ${\tau}^{\pm}$ ${\to}$
  $K^{{\ast}{\pm}}{\nu}_{\tau}$ decays.
  Using Eq.(\ref{eq:tau-decay-width}) and experimental data
  on branching ratio ${\cal B}({\tau}{\to}K^{\ast}{\nu})$
  $=$ $1.20(7)$ \% \cite{pdg2020}, one can obtain the decay constant
  $f_{K^{\ast}}^{\rm exp}$ $=$ $202.5^{+6.5}_{-6.7}$ MeV.
  The value of $f_{K^{\ast}}^{\rm exp}$ is much less than that
  of LQCD results, and will be used in our calculation.

  For the $K^{{\ast}-}$ ${\to}$ ${\ell}^{-}\bar{\nu}_{\ell}$ decays,
  the SM expectations on the partial decay widths and branching
  ratios are,
   \begin{eqnarray}
  {\Gamma}(K^{{\ast}-}{\to}e^{-}\bar{\nu}_{e})
   &=& 5.5{\pm}0.4\, {\mu}\,{\rm eV}
   \label{eq:width-kv-e}, \\
  {\Gamma}(K^{{\ast}-}{\to}{\mu}^{-}\bar{\nu}_{\mu})
   &=& 5.3{\pm}0.4\, {\mu}\,{\rm eV}
   \label{eq:width-kv-mu}, \\
   {\cal B}(K^{{\ast}-}{\to}e^{-}\bar{\nu}_{e})
   &=& (1.2{\pm}0.1){\times}10^{-13}
   \label{eq:br-kv-e}, \\
  {\cal B}(K^{{\ast}-}{\to}{\mu}^{-}\bar{\nu}_{\mu})
   &=& (1.2{\pm}0.1){\times}10^{-13}
   \label{eq:br-kv-mu}.
   \end{eqnarray}
  The decay width ${\Gamma}_{K^{\ast}}$ $=$ $46.2{\pm}1.3$ MeV
  \cite{pdg2020} is used in our calculation.
  It is apparent that more than $10^{14}$ $K^{{\ast}{\pm}}$ events
  are the minimum requirement for experimentally studying
  the $K^{{\ast}-}$ ${\to}$ ${\ell}^{-}\bar{\nu}_{\ell}$ decays.

  Based on the $U$-spin symmetry, the production mechanism of
  the $K^{{\ast}{\pm}}$ mesons in electron-position collisions
  is similar to that of the ${\rho}^{\pm}$ mesons.
  An educated guess is that the cross section
  ${\sigma}(e^{+}e^{-}{\to}K^{{\ast}+}K^{{\ast}-})$ ${\sim}$
  $20$ fb and $230$ fb near $\sqrt{s}$ ${\sim}$ $m_{{\Upsilon}(4S)}$
  and $m_{J/{\psi}}$, respectively.
  The branching ratios of $J/{\psi}$ decays are \cite{pdg2020},
   \begin{eqnarray}
  {\cal B}(J/{\psi}{\to}K^{{\ast}+}K^{{\ast}-})
   &=& (1.00^{+0.22}_{-0.40}){\times}10^{-3}
   \label{eq:br-j/psi-kstar-pair}, \\
  {\cal B}(J/{\psi}{\to}K^{{\ast}{\pm}}K^{\mp})
   &=& (6.0^{+0.8}_{-1.0}){\times}10^{-3}
   \label{eq:br-j/psi-kstar-k}, \\
  {\cal B}(J/{\psi}{\to}K^{{\ast}{\pm}}K^{\mp}{\pi}^{0})
   &=& (4.1{\pm}1.3){\times}10^{-3}
   \label{eq:br-j/psi-kstar-k-piz}, \\
  {\cal B}(J/{\psi}{\to}K^{{\ast}{\pm}}K_{S}^{0}{\pi}^{\mp})
   &=& (2.0{\pm}0.5){\times}10^{-3}
   \label{eq:br-j/psi-kstar-ks-pi}, \\
  {\cal B}(J/{\psi}{\to}K^{{\ast}{\pm}}K_{2}^{\ast}(1430)^{\mp})
   &=& (3.4{\pm}2.9){\times}10^{-3}
   \label{eq:br-j/psi-kstar-k1430}, \\
  {\cal B}(J/{\psi}{\to}K^{{\ast}{\pm}}K^{\ast}(700)^{\mp})
   &=& (1.0^{+1.0}_{-0.6}){\times}10^{-3}
   \label{eq:br-j/psi-kstar-k1430}.
   \end{eqnarray}
  It is approximately estimated that
  ${\cal B}(J/{\psi}{\to}K^{{\ast}{\pm}}X^{\mp})$ ${\sim}$
  $1.8$ \%.
  Hence, the experiemtal data on the $K^{{\ast}{\pm}}$
  mesons at the $e^{+}e^{-}$ collisions,
  which would be available by either the prompt
  $K^{{\ast}+}K^{{\ast}-}$ pair production at SuperKEKB and
  SCTF experiments or the production via $10^{13}$ $J/{\psi}$
  decay at SCTF, are far from sufficient for investigating
  the $K^{{\ast}-}$ ${\to}$ ${\ell}^{-}\bar{\nu}_{\ell}$ decays.
  If we assume that the inclusive cross section of prompt
  $K^{{\ast}{\pm}}$ meson production in $pp$ collisions at the
  center-of-mass energy of $13$ TeV is similar to that of
  ${\rho}^{\pm}$ mesons, about ${\cal O}(10\,{\rm mb})$,
  there would be some $3{\times}10^{15}$ $K^{{\ast}{\pm}}$
  events to be available with an integrated
  luminosity of $300$ ${\rm fb}^{-1}$ at LHCb, which correspond
  to about ${\cal O}(10^{2})$ events of the $K^{{\ast}-}$ ${\to}$
  ${\ell}^{-}\bar{\nu}_{\ell}$ decays.
  It should be some glimmer of hope for observation and
  scrutinies of the $K^{{\ast}-}$ ${\to}$
  ${\ell}^{-}\bar{\nu}_{\ell}$ decays at hadron-hadron
  collisions in the future, particularly at the planning HE-HLC.

  \section{$D_{d}^{{\ast}-}$ ${\to}$ ${\ell}^{-}\bar{\nu}_{\ell}$ decays}
  \label{sec-cd}
  The mass of $D_{d}^{\ast}$ mesons, $m_{D_{d}^{\ast}}$ $=$
  $2010.26(5)$ MeV, are just above the threshold of $D{\pi}$ pair.
  The $D_{d}^{\ast}$ meson decays via the strong interactions
  are dominant, and the ratio of branching ratios \cite{pdg2020},
  ${\cal B}(D_{d}^{{\ast}{\pm}}{\to}D_{d}^{\pm}{\pi}^{0})/
   {\cal B}(D_{d}^{{\ast}{\pm}}{\to}D_{u}{\pi}^{\pm})$ $=$
  $30.7(5)\,\%/67.7(5)\,\%$ ${\sim}$ $1/2$, basically agrees
  with the relations of isospin symmetry.
  It should be pointed out that the $D_{d}^{\ast}$ strong decays
  are highly suppressed by the compact phase spaces because of
  $m_{D_{d}^{\ast}}$ $-$ $m_{D}$ $-$ $m_{\pi}$ $<$ 6 MeV.
  The branching ratio of the magnetic dipole transition
  is small, ${\cal B}(D_{d}^{\ast}{\to}D_{d}{\gamma})$
  $=$ $1.6(4)\,\%$ \cite{pdg2020}.
  Hence, the decay width of $D_{d}^{\ast}$ mesons is narrow,
  ${\Gamma}_{D_{d}^{\ast}}$ $=$ $83.4{\pm}1.8$ keV \cite{pdg2020}.
  From the $D_{d}^{{\ast}-}$ ${\to}$ ${\ell}^{-}\bar{\nu}_{\ell}$ decays,
  the parameter ${\vert}V_{cd}{\vert}\,f_{D_{d}^{\ast}}$ is expected
  to be experimentally determined.

  \begin{figure}[ht]
  \includegraphics[width=0.55\textwidth]{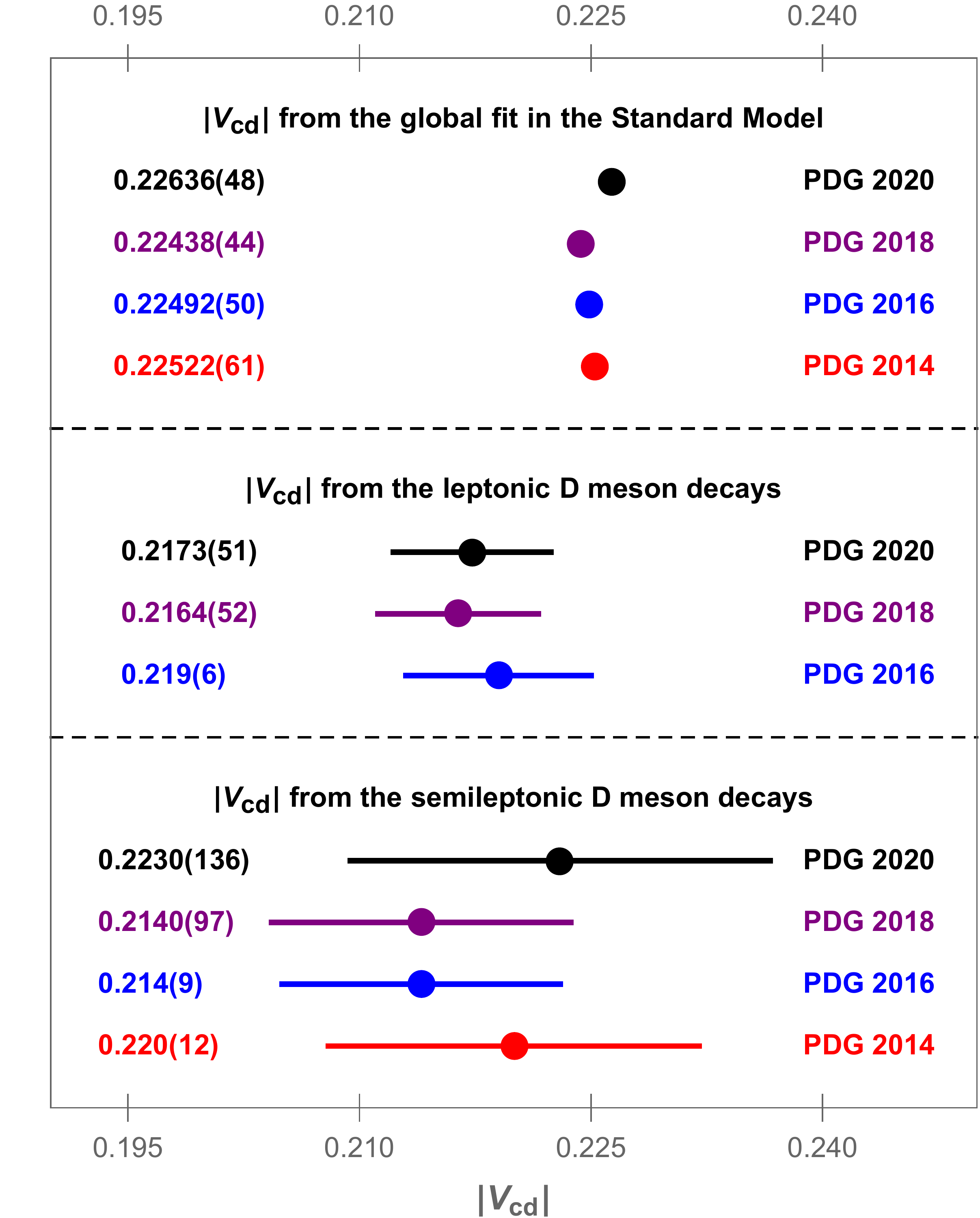}
  \caption{The values of the CKM element ${\vert}V_{cd}{\vert}$
  from PDG.}
  \label{fig:vcd}
  \end{figure}
   \begin{table}[ht]
   \caption{The theoretical values of decay constant $f_{D_{d}^{\ast}}$
   (in the unit of MeV), where the legends are the same as those in
   Table \ref{tab:rho-decay-constant}, NRQM is an abbreviation for
   nonrelativistic quark model.}
   \label{tab:cdstar-decay-constant}
   \begin{ruledtabular}
   \begin{tabular}{lccccc}
  NRQM & $223^{+23}_{-19}$ \cite{PhysRevD.71.113006}
       & $307$ \cite{epl.115.21002} \footnotemark[1]
       & $253$ \cite{epl.115.21002} \footnotemark[2]
       & $353.8$ \cite{epjp.132.80} \footnotemark[1]
       & $290.3$ \cite{epjp.132.80} \footnotemark[2] \\
  NRQM & $391$ \cite{plb.635.93}
       & $290$ \cite{ahep.2018.7032041} \footnotemark[1]
       & $210$ \cite{ahep.2018.7032041} \footnotemark[2]
       & $332$ \cite{mpla.17.803} \\
  RQM  & $310$ \cite{plb.635.93}
       & $315$ \cite{mpla.17.803}
       & $327{\pm}13$ \cite{PhysRevD.55.6944} \footnotemark[3]
       & $252{\pm}10$ \cite{PhysRevD.55.6944} \footnotemark[4] \\
  LFQM & $254$ \cite{PhysRevD.75.073016} \footnotemark[5]
       & $228$ \cite{PhysRevD.75.073016} \footnotemark[6]
       & $259.6{\pm}14.6$ \cite{PhysRevD.81.114024} \footnotemark[7]
       & $306.3^{+18.2}_{-17.7}$ \cite{PhysRevD.81.114024} \footnotemark[8]
       & $253{\pm}7$ \cite{PhysRevD.98.114018} \\
  LFQM & $230$ \cite{PhysRevC.92.055203} \footnotemark[9]
       & $226.6^{+~5.9}_{-10.2}$ \cite{cpc.42.073102} \footnotemark[10]
       & $230.1{\pm}6.2$ \cite{cpc.42.073102} \footnotemark[11]
       & $245^{+35}_{-34}$ \cite{jpg.39.025005}
       & $230{\pm}29$ \cite{PhysRevD.100.014026} \\
  LFQM & $252.0^{+13.8}_{-11.6}$ \cite{epjc.76.313} \footnotemark[7]
       & $264.9^{+10.2}_{-~9.5}$ \cite{epjc.76.313} \footnotemark[8]
       & $272$ \cite{epjp.133.134} \footnotemark[12]
       & $260$ \cite{epjp.133.134} \footnotemark[13]
       & $269$ \cite{epjp.133.134} \footnotemark[14] \\
  LQCD  & $245{\pm}20$ \cite{PhysRevD.60.074501}
        & $234{\pm}26$ \cite{npb.619.507} \footnotemark[15]
        & $278{\pm}16$ \cite{jhep.1202.042}
        & $223.5{\pm}8.7$ \cite{PhysRevD.96.034524}
        & $234(6)$ \cite{cpc.45.023109} \\
  SR    & $242^{+20}_{-12}$ \cite{PhysRevD.88.014015}
        & $263{\pm}21$ \cite{epjc.75.427}
        & $252.2{\pm}22.7$ \cite {plb.735.12}
        & $250{\pm}11$ \cite{ijmpa.30.1550116} \\
  other & $186$ \cite{epjp.133.134} \footnotemark[16]
        & $273{\pm}13$ \cite{PhysRevD.75.116001}
        & $341{\pm}23$ \cite{plb.633.492}
        & $237$ \cite{PhysRevD.58.014007} \\
  \end{tabular}
  \end{ruledtabular}
  \footnotetext[1]{Without QCD radiative corrections.}
  \footnotetext[2]{With QCD radiative corrections.}
  \footnotetext[3]{With constituent quark masses for the light quarks $m_{s}$ and $m_{d}$.}
  \footnotetext[4]{With current quark masses for the light quarks $m_{s}$ and $m_{d}$.}
  \footnotetext[5]{With Coulomb plus linear potential model.}
  \footnotetext[6]{With Coulomb plus harmonic oscillator potential model.}
  \footnotetext[7]{With the Gaussian type wave functions.}
  \footnotetext[8]{With the power-law type wave functions.}
  \footnotetext[9]{With Coulomb plus linear potential model.}
  \footnotetext[10]{With a dilation parameter ${\kappa}$ $=$ $0.54$ GeV.}
  \footnotetext[11]{With a dilation parameter ${\kappa}$ $=$ $0.68$ GeV.}
  \footnotetext[12]{With Martin potential model \cite{plb.93.338}.}
  \footnotetext[13]{With Cornell potential model \cite{zpc.33.135}.}
  \footnotetext[14]{With logarithmic potential model \cite{plb.71.153}.}
  \footnotetext[15]{With $m_{D^{\ast}}/f_{D^{\ast}}$ $=$ $8.6{\pm}0.3^{+0.5}_{-0.9}$
                    \cite{npb.619.507} and $m_{D_{d}^{\ast}}$ $=$ $2010.26(5)$ MeV
                    \cite{pdg2020}.}
  \footnotetext[16]{With harmonic plus Yukawa potential model \cite{epjp.132.80}.}
  \end{table}

  Currently, the precise values of the CKM element ${\vert}V_{cd}{\vert}$
  comes mainly from the leptonic and semileptonic $D$ meson decays
  \cite{pdg2020}, as illustrated in Fig. \ref{fig:vcd}.
  Because of the decay width of Eq.(\ref{eq:decay-width-pseudoscalar})
  being proportional to $m_{\ell}^{2}$, the $D^{-}$ ${\to}$
  $e^{-}\bar{\nu}_{e}$ decay is helicity suppressed.
  And the $D^{-}$ ${\to}$ ${\tau}^{-}\bar{\nu}_{\tau}$ decay suffers
  from the complications caused by the additional neutrino in
  ${\tau}$ decays. The $D^{-}$ ${\to}$ ${\mu}^{-}\bar{\nu}_{\mu}$
  decay is the most favorable mode for experimental measurement.
  For the values of ${\vert}V_{cd}{\vert}$ from the purely leptonic
  decay $D^{-}$ ${\to}$ ${\mu}^{-}\bar{\nu}$, the experimentally
  statistical uncertainties are dominant uncertainties.
  For the values of ${\vert}V_{cd}{\vert}$ from the semileptonic
  $D$ meson decays, the theoretical uncertainties from the form
  factor controlled by nonperturbative dynamics are dominant
  uncertainties.
  It is clearly seen from Fig. \ref{fig:vcd} that the experimental
  uncertainties have not decreased significantly recently.
  Besides, ${\vert}V_{cd}{\vert}$ can also be determined from
  the neutrino-induced charm production data \cite{pdg2020},
  but the relevant experimental data have not been updated
  after the measurements given by the CHARM-II Collaboration
  in 1999 \cite{epjc.11.19}.
  According to the Wolfenstein parameterization of the CKM
  matrix, there is an approximate relation between its elements
  ${\vert}V_{cd}{\vert}$ $=$ ${\vert}V_{us}{\vert}$ $=$
  ${\lambda}$ up to ${\cal O}({\lambda}^{4})$.
  However, the measurement precision of the CKM element
  ${\vert}V_{cd}{\vert}$ from both leptonic and
  semileptonic $D$ meson decays is generally about an order
  of magnitude smaller than that of ${\vert}V_{us}{\vert}$
  from leptonic and semileptonic $K$ meson decays
  for the moment.
  The most precise values are from the global fit in SM,
  ${\vert}V_{cd}{\vert}$ $=$ $0.22636(48)$ \cite{pdg2020}
  with uncertainties ${\sim}$ $0.2\,\%$.

  The information about the decay constant $f_{D_{d}^{\ast}}$
  has not yet been obtained experimentally by now.
  Some theoretical results on $f_{D_{d}^{\ast}}$ are listed
  in Table \ref{tab:cdstar-decay-constant}.
  The theoretical discrepancies among various methods are
  obvious.
  In our calculation, as a conservative estimate, we will
  take the recent value $f_{D_{d}^{\ast}}$ $=$ $230{\pm}29$
  MeV \cite{PhysRevD.100.014026} from the light front quark model,
  which agrees basically with the values $f_{D_{d}^{\ast}}$
  $=$ $234{\pm}6$ MeV \cite{cpc.45.023109} from the recent
  lattice QCD simulation.

  After some simple computation with Eq.(\ref{eq:decay-width--vector}),
  we obtain the partial decay widths and branching ratios for the
  $D_{d}^{{\ast}-}$ ${\to}$ ${\ell}^{-}\bar{\nu}_{\ell}$
  decays as follows.
   \begin{eqnarray}
  {\Gamma}(D_{d}^{{\ast}-}{\to}{\ell}^{-}\bar{\nu}_{\ell})
   &=& 79^{+22}_{-19}\, {\mu}\,{\rm eV}, \quad\text{for}\
   {\ell}\, =\, e,\,{\mu}
   \label{eq:width-cd-e-mu}, \\
  {\Gamma}(D_{d}^{{\ast}-}{\to}{\tau}^{-}\bar{\nu}_{\tau})
   &=& 5{\pm}1\, {\mu}\,{\rm eV}
   \label{eq:width-cd-tau}, \\
   {\cal B}(D_{d}^{{\ast}-}{\to}{\ell}^{-}\bar{\nu}_{\ell})
   &=& (9.5^{+2.9}_{-2.4}){\times}10^{-10}, \quad\text{for}\
   {\ell}\, =\, e,\,{\mu}
   \label{eq:br-cd-e-mu}, \\
   {\cal B}(D_{d}^{{\ast}-}{\to}{\tau}^{-}\bar{\nu}_{\tau})
   &=& (0.6{\pm}0.2){\times}10^{-10}
   \label{eq:br-cd-tau}.
   \end{eqnarray}
  These branching ratios are consistent with those of
  Ref. \cite{ctp.67.655} if the different values of decay
  constants $f_{D_{d}^{\ast}}$ are considered.
  The relatively large uncertainties of 
  branching ratios come from the uncertainties of mass
  $m_{D_{d}^{\ast}}$, width ${\Gamma}_{D_{d}^{\ast}}$,
  decay constant $f_{D_{d}^{\ast}}$ and the CKM element
  ${\vert}V_{cd}{\vert}$.
  To experimentally study the $D_{d}^{{\ast}-}$ ${\to}$
  ${\ell}^{-}\bar{\nu}_{\ell}$ decays,
  more than $10^{11}$ $D_{d}^{\ast}$ events are needed.
  Due to the short lifetime of lepton ${\tau}^{\pm}$
  and the lepton number conservation in ${\tau}^{\pm}$ decays,
  additional neutrinos will make the measurement of the
  $D_{d}^{{\ast}-}$ ${\to}$ ${\tau}^{-}\bar{\nu}_{\tau}$
  decay to have a poor reconstruction efficiency and
  to be very challenging.
  Perhaps some $10^{12}$ or more $D_{d}^{\ast}$ events
  are necessarily required to study the $D_{d}^{{\ast}-}$
  ${\to}$ ${\tau}^{-}\bar{\nu}_{\tau}$ decay.

  \begin{figure}[ht]
  \includegraphics[width=0.4\textwidth]{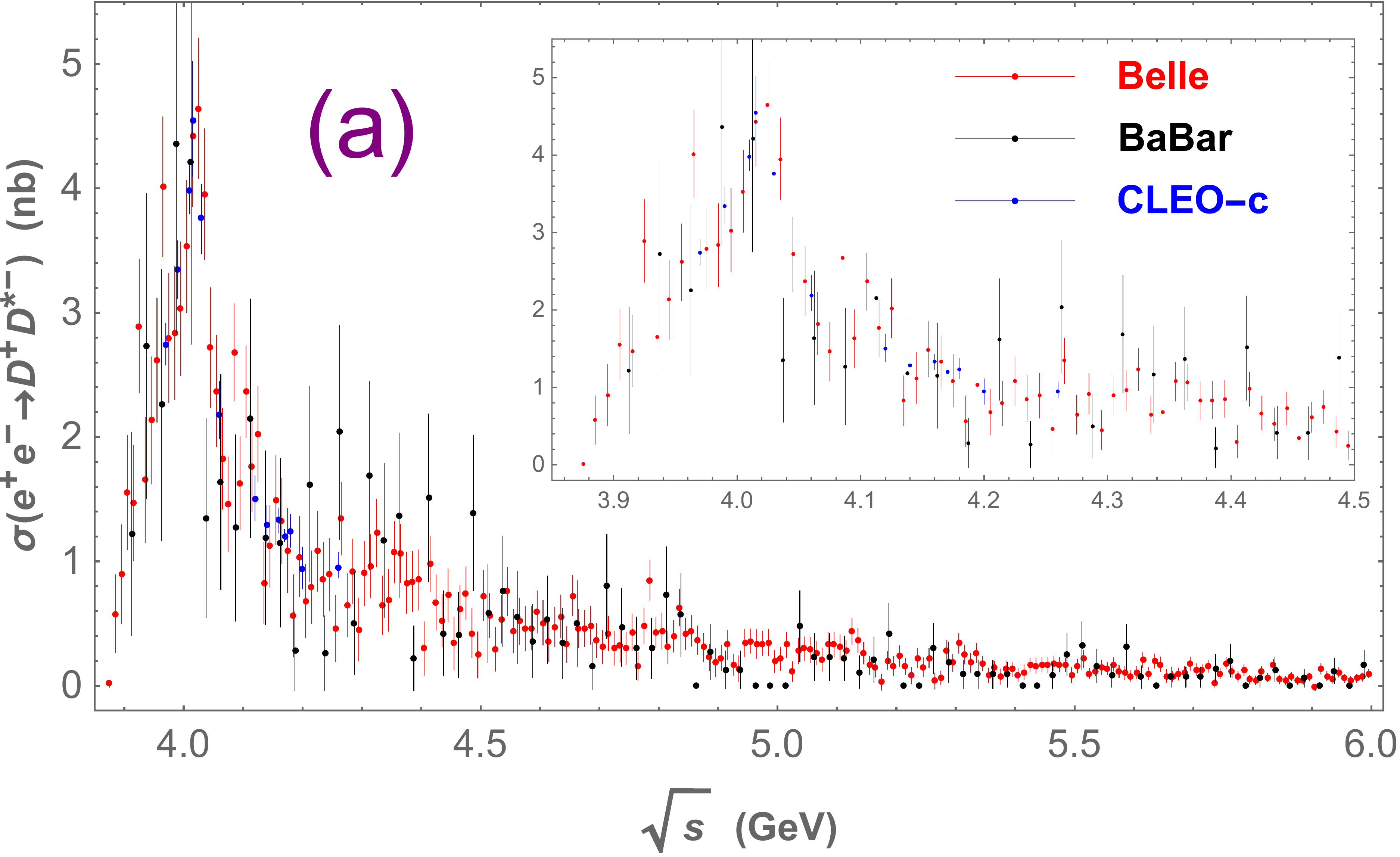}\quad
  \includegraphics[width=0.4\textwidth]{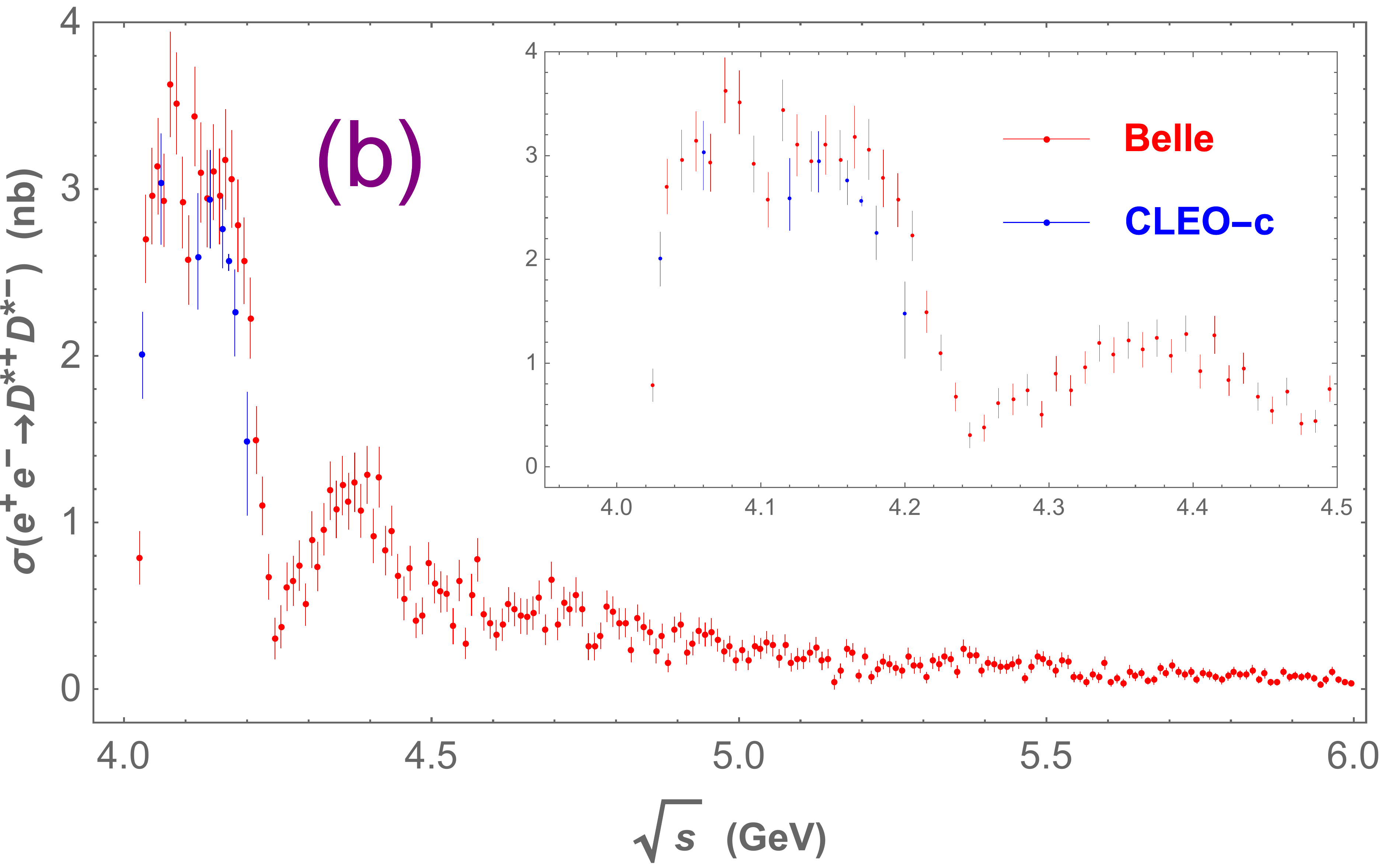}
  \caption{The exclusive cross sections (in the unit of nb)
  as functions of $\sqrt{s}$ (in the unit of GeV) for
  $e^{+}e^{-}$ ${\to}$ $D^{+}D^{{\ast}-}$
  in (a) and $e^{+}e^{-}$ ${\to}$ $D^{{\ast}+}D^{{\ast}-}$
  in (b).
  The Belle, BaBar and CLEO-c data are from Ref.
  \cite{PhysRevD.97.012002}, Ref. \cite{PhysRevD.79.092001}
  and Ref. \cite{cpc.42.043002}, respectively.}
  \label{fig:cs-dv}
  \end{figure}

  Above the open charm production threshold, there are several
  charmonium resonances and charmonium-like structures decaying
  predominantly into pairs of charmed meson final states.
  The studies of Belle \cite{PhysRevD.97.012002},
  BaBar \cite{PhysRevD.79.092001} and
  CLEO-c \cite{PhysRevD.80.072001,cpc.42.043002}
  collaborations have shown that
  there is a sharply peaked $D_{d}^{+}D_{d}^{{\ast}-}$ structure
  and a broad $D_{d}^{{\ast}+}D_{d}^{{\ast}-}$ plateau
  just above threshold, as illustrated in Fig.~\ref{fig:cs-dv}.
  Assuming the exclusive cross sections near threshold
  ${\sigma}(e^{+}e^{-}{\to}D_{d}^{+}D_{d}^{{\ast}-})$ ${\sim}$
  $4\,{\rm nb}$ and
  ${\sigma}(e^{+}e^{-}{\to}D_{d}^{{\ast}+}D_{d}^{{\ast}-})$ ${\sim}$
  $3\,{\rm nb}$, there will be about $10^{11}$ $D_{d}^{{\ast}{\pm}}$
  events corresponding to the total integrated luminosity of
  $10$ ${\rm ab}^{-1}$ at future STCF, and about $5{\times}10^{11}$
  $D_{d}^{{\ast}{\pm}}$
  events corresponding to a data sample of $50$ ${\rm ab}^{-1}$
  at SuperKEKB.
  In addition, about $10^{12}$ $Z$ bosons will be produced on the
  on the schedule of the large international scientific project
  of Circular Electron Positron Collider (CEPC) \cite{cepc}
  and $10^{13}$ $Z$ bosons at Future Circular $e^{+}e^{-}$
  Collider (FCC-ee) \cite{fcc}.
  Considering the branching ratio ${\cal B}(Z{\to}D^{{\ast}{\pm}}X)$ $=$
  $(11.4{\pm}1.3)\,\%$ \cite{pdg2020}, the $Z$ boson decays will
  yield more than $10^{11}$ $D_{d}^{{\ast}{\pm}}$ events at the
  tera-Z factories.
  So the $D_{d}^{{\ast}-}$ ${\to}$ ${\ell}^{-}\bar{\nu}_{\ell}$
  decays could be investigated at Belle-II, SCTF, CEPC and
  FCC-ee experiments.

  In hadron-hadron collisions, the inclusive cross sections for
  the $c\bar{c}$ pair and $D^{{\ast}+}$ meson production are
  measured to be ${\sigma}(pp{\to}c\bar{c}X)$ $=$
  $2369{\pm}3{\pm}152{\pm}118$ ${\mu}$b and ${\sigma}(pp{\to}D^{{\ast}+}X)$ $=$
  $784{\pm}4{\pm}87{\pm}118$ ${\mu}$b at the center-of-mass
  energy of $\sqrt{s}$ $=$ $13$ TeV by the LHCb group, with the
  transverse momentum $p_{T}$ within the range of $1\,{\rm GeV}$
  $<$ $p_{T}$ $<$ $8\,{\rm GeV}$ \cite{JHEP.2016.03.159}.
  Some $2{\times}10^{14}$ $D_{d}^{{\ast}{\pm}}$ events could be
  accumulated with the integrated luminosity $300\,{\rm fb}^{-1}$
  at LHCb.
  The total cross sections of charm and $D^{{\ast}+}$
  production measured at $\sqrt{s}$ $=$ $7$ TeV by the ALICE group
  are ${\sigma}_{c\bar{c}}^{\rm tot}$ ${\simeq}$ $8.5$ mb
  and ${\sigma}_{D^{{\ast}+}}^{\rm tot}$ ${\simeq}$ $2.11$ mb,
  respectively \cite{JHEP.2012.07.191}.
  The total cross sections of charm production measured at
  $\sqrt{s}$ $=$ $7$ TeV by the ATLAS group are
  ${\sigma}_{c\bar{c}}^{\rm tot}$ ${\simeq}$ $8.6$ mb
  \cite{npb.907.717}.
  The $D^{{\ast}+}$ production cross section at ATLAS
  should be very close to that at ALICE based on an educated
  guess.
  In addition, the $D^{\ast}$ meson can also produced from
  $b$ decays with the fragmentation fraction about
  $f(b{\to}D_{d}^{\ast})$ ${\simeq}$ $23\,\%$ \cite{epjc.75.19}.
  The $b$-quark production cross sections at $\sqrt{s}$ $=$ 13 TeV
  determined by LHCb and ALICE are about ${\sigma}(pp{\to}b\bar{b}X)$
  ${\simeq}$ $495$ ${\mu}$b \cite{JHEP.2015.10.172} and
  $541$ ${\mu}$b \cite{2108.02523}, respectively.
  So more than $10^{13}$ $D_{d}^{{\ast}{\pm}}$ events from $b$
  decays could be accumulated with the integrated luminosity
  $300\,{\rm fb}^{-1}$ at LHCb.
  All in all, the large cross section of $D^{\ast}$ meson plus
  the high luminosity at hadron-hadron collisions result in
  the abundant $D_{d}^{{\ast}{\pm}}$ events, and make the carefully
  experimental study of the $D_{d}^{{\ast}-}$ ${\to}$
  $e^{-}\bar{\nu}_{e}$, ${\mu}^{-}\bar{\nu}_{\mu}$ decays, even the
  $D_{d}^{{\ast}-}$ ${\to}$ ${\tau}^{-}\bar{\nu}_{\tau}$ decay,
  to be possible and practicable.

  \section{$D_{s}^{{\ast}-}$ ${\to}$ ${\ell}^{-}\bar{\nu}_{\ell}$ decays}
  \label{sec-cs}
  The $D_{s}^{{\ast}{\pm}}$ mesons have explicitly nonzero quantum number
  of electric charges, charm and strange, $Q$ $=$ $C$ $=$ $S$ $=$ ${\pm}1$.
  Considering the conservation of the charm and strange quantum number
  in the strong and electromagnetic interactions,
  and the mass of $D_{s}^{\ast}$ mesons, $m_{D_{s}^{\ast}}$
  $=$ $2112.2(4)$ MeV \cite{pdg2020}, being just above the threshold
  of $D_{s}{\pi}$ pair but below the threshold of $DK$ pair,
  the $D_{s}^{\ast}$ ${\to}$ $D_{s}{\pi}$ decays are the only allowable
  hadronic decay modes.
  However, the $D_{s}^{\ast}$ ${\to}$ $D_{s}{\pi}$ decays are highly
  suppressed due to four factors:
  (1) from the dynamical view, the $D_{s}^{\ast}$ ${\to}$ $D_{s}{\pi}$
  decays are induced by the the electromagnetic interactions rather
  than the strong interactions because of the isospin non-conservation
  between the initial and final states,
  (2) from the perspective of the conservation of angular momentum,
  the orbital angular momentum of final states should be $L$ $=$ $1$,
  so the $D_{s}^{\ast}$ ${\to}$ $D_{s}{\pi}$ decays are induced by
  the contributions of the $P$-wave amplitudes,
  (3) from the phenomenological view, the $D_{s}^{\ast}$ ${\to}$
  $D_{s}{\pi}$ decays are suppressed by the the Okubo-Zweig-Iizuka
  rules \cite{zweig,ozi-o,ozi-i} because the quark lines of pion
  disconnect from those of the $D_{s}^{\ast}D_{s}$ system,
  (4) from the kinematic view, the phase spaces of final states are
  very compact because of $m_{D_{s}^{\ast}}$ $-$ $m_{D_{s}}$ $-$
  $m_{{\pi}^{0}}$ ${\sim}$ 9 MeV.
  Hence, the branching ratio for the hadronic decay is very small
  ${\cal B}(D_{s}^{\ast}{\to}D_{s}{\pi})$ $=$ $5.8(7)\,\%$
  \cite{pdg2020}. And
  the branching ratio of the electromagnetic radiative decay is
  dominant, ${\cal B}(D_{s}^{\ast}{\to}D_{s}{\gamma})$ $=$
  $93.5(7)\,\%$ \cite{pdg2020}.
  Except for the $D_{s}{\pi}$, $D_{s}{\gamma}$ and $D_{s}e^{+}e^{-}$
  final states, other decay modes of the $D_{s}^{\ast}$ mesons
  have not yet been observed \cite{pdg2020}.
  The $D_{s}^{{\ast}-}$ ${\to}$ ${\ell}^{-}\bar{\nu}_{\ell}$
  weak decays are favored by the CKM element ${\vert}V_{cs}{\vert}$.
  The information about the ${\vert}V_{cs}{\vert}f_{D_{s}^{\ast}}$
  can be experimentally obtained from the $D_{s}^{{\ast}-}$ ${\to}$
  ${\ell}^{-}\bar{\nu}_{\ell}$ decays.

  \begin{figure}[ht]
  \includegraphics[width=0.55\textwidth]{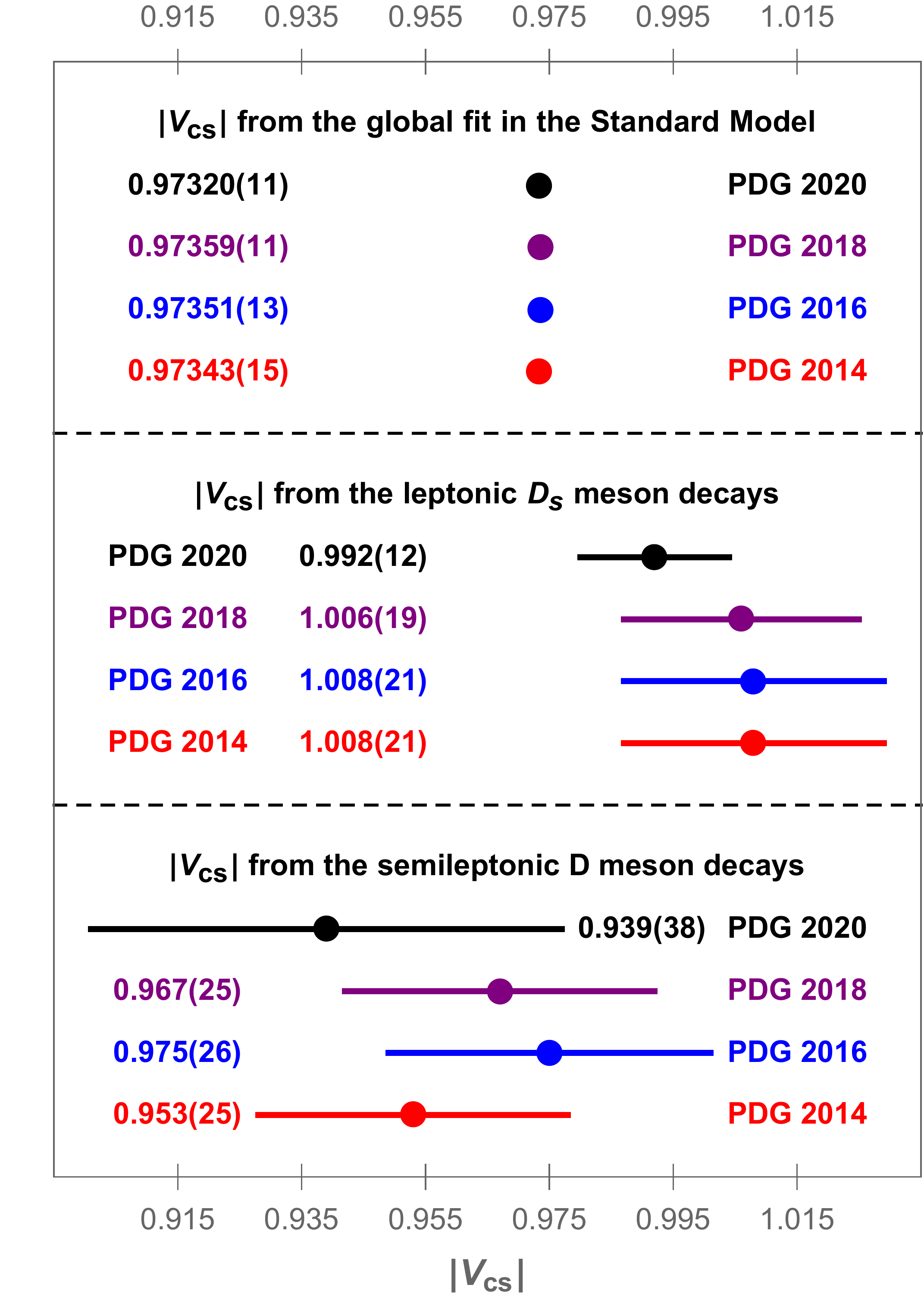}
  \caption{The values of the CKM element ${\vert}V_{cs}{\vert}$
  from PDG.}
  \label{fig:vcs}
  \end{figure}
   \begin{table}[ht]
   \caption{The theoretical values of decay constant $f_{D_{s}^{\ast}}$
   (in the unit of MeV), where the legends including the footnotes
   are the same as those in \ref{tab:cdstar-decay-constant}.}
   \label{tab:csstar-decay-constant}
   \begin{ruledtabular}
   \begin{tabular}{lccccc}
  NRQM & $326^{+21}_{-17}$ \cite{PhysRevD.71.113006}
       & $344$ \cite{epl.115.21002} \footnotemark[1]
       & $275$ \cite{epl.115.21002} \footnotemark[2]
       & $382.1$ \cite{epjp.132.80} \footnotemark[1]
       & $303.5$ \cite{epjp.132.80} \footnotemark[2] \\
  NRQM & $447$ \cite{plb.635.93}
       & $310$ \cite{ahep.2018.7032041} \footnotemark[1]
       & $212$ \cite{ahep.2018.7032041} \footnotemark[2]
       & $384$ \cite{mpla.17.803} \\
  RQM  & $315$ \cite{plb.635.93}
       & $335$ \cite{mpla.17.803}
       & $362{\pm}15$ \cite{PhysRevD.55.6944} \footnotemark[3]
       & $288{\pm}11$ \cite{PhysRevD.55.6944} \footnotemark[4]
       & $272$ \cite{PhysRevD.96.016017} \\
  LFQM & $290$ \cite{PhysRevD.75.073016} \footnotemark[5]
       & $268$ \cite{PhysRevD.75.073016} \footnotemark[6]
       & $338.7{\pm}29.7$ \cite{PhysRevD.81.114024} \footnotemark[7]
       & $391.0{\pm}28.9$ \cite{PhysRevD.81.114024} \footnotemark[8]
       & $314{\pm}6$ \cite{PhysRevD.98.114018} \\
  LFQM & $260$ \cite{PhysRevC.92.055203} \footnotemark[9]
       & $254.7^{+6.3}_{-6.7}$ \cite{cpc.42.073102} \footnotemark[10]
       & $289.7^{+6.3}_{-4.5}$ \cite{cpc.42.073102} \footnotemark[11]
       & $272^{+39}_{-38}$ \cite{jpg.39.025005}
       & $253{\pm}32$ \cite{PhysRevD.100.014026} \\
  LFQM & $318.3^{+15.3}_{-12.6}$ \cite{epjc.76.313} \footnotemark[7]
       & $330.9^{+9.9}_{-9.0}$ \cite{epjc.76.313} \footnotemark[8]
       & $303$ \cite{epjp.133.134} \footnotemark[12]
       & $291$ \cite{epjp.133.134} \footnotemark[13]
       & $302$ \cite{epjp.133.134} \footnotemark[14] \\
  LQCD  & $272{\pm}16^{+~3}_{-20}$ \cite{PhysRevD.60.074501}
        & $254{\pm}17$ \cite{npb.619.507} \footnotemark[15]
        & $311{\pm}9$ \cite{jhep.1202.042} \\
  LQCD  & $268.8{\pm}6.6$ \cite{PhysRevD.96.034524}
        & $274{\pm}6$ \cite{PhysRevLett.112.212002}
        & $274{\pm}7$ \cite{cpc.45.023109} \\
  SR    & $293^{+19}_{-14}$ \cite{PhysRevD.88.014015}
        & $308{\pm}21$ \cite{epjc.75.427}
        & $305.5{\pm}27.3$ \cite {plb.735.12}
        & $270{\pm}19$ \cite{ijmpa.30.1550116} \\
  other & $240$ \cite{epjp.133.134} \footnotemark[16]
        & $307{\pm}18$ \cite{PhysRevD.75.116001}
        & $375{\pm}24$ \cite{plb.633.492}
        & $242$ \cite{PhysRevD.58.014007} \\
  \end{tabular}
  \end{ruledtabular}
  \footnotetext[1]{Without QCD radiative corrections.}
  \footnotetext[2]{With QCD radiative corrections.}
  \footnotetext[3]{With constituent quark masses for the light quarks $m_{s}$ and $m_{d}$.}
  \footnotetext[4]{With current quark masses for the light quarks $m_{s}$ and $m_{d}$.}
  \footnotetext[5]{With Coulomb plus linear potential model.}
  \footnotetext[6]{With Coulomb plus harmonic oscillator potential model.}
  \footnotetext[7]{With the Gaussian type wave functions.}
  \footnotetext[8]{With the power-law type wave functions.}
  \footnotetext[9]{With Coulomb plus linear potential model.}
  \footnotetext[10]{With a dilation parameter ${\kappa}$ $=$ $0.54$ GeV.}
  \footnotetext[11]{With a dilation parameter ${\kappa}$ $=$ $0.68$ GeV.}
  \footnotetext[12]{With Martin potential model \cite{plb.93.338}.}
  \footnotetext[13]{With Cornell potential model \cite{zpc.33.135}.}
  \footnotetext[14]{With logarithmic potential model \cite{plb.71.153}.}
  \footnotetext[15]{With $m_{D_{s}^{\ast}}/f_{D_{s}^{\ast}}$ $=$ $8.3{\pm}0.2{\pm}0.5$
                    \cite{npb.619.507} and $m_{D_{s}^{\ast}}$ $=$ $2112.2(4)$ MeV
                    \cite{pdg2020}.}
  \footnotetext[16]{With harmonic plus Yukawa potential model \cite{epjp.132.80}.}
  \end{table}

  The direct determinations of the CKM element ${\vert}V_{cs}{\vert}$
  come mainly from leptonic $D_{s}$ decays and semileptonic $D$ decays,
  as shown in Fig.~\ref{fig:vcs}.
  The uncertainties of ${\vert}V_{cs}{\vert}$ from the $D_{s}$ leptonic
  decays, about $1\,\%$, are dominated by the experimental uncertainties.
  The uncertainties of ${\vert}V_{cs}{\vert}$ from the $D$ semileptonic
  decays, about $4\,\%$, are dominated by the theoretical
  calculations of the form factors.
  It is wroth noting that the recent CKM element ${\vert}V_{cs}{\vert}$
  determined by the BES-III group from the $D_{s}^{+}$ ${\to}$
  ${\mu}^{+}{\nu}_{\mu}$ and $D_{s}^{+}$ ${\to}$ ${\tau}^{+}{\nu}_{\tau}$
  decays based on available $6.32$ ${\rm fb}^{-1}$ data is
  ${\vert}V_{cs}{\vert}$ $=$ $0.978{\pm}0.009{\pm}0.014$
  \cite{PhysRevD.104.052009}, where the systematic (second) uncertainties
  has outweighed the statistical (first) one.
  This value is very close to the precise result from the global fit,
  ${\vert}V_{cs}{\vert}$ $=$ $0.97320(11)$
  \cite{pdg2020} that will be used in this paper.

  By now, a relatively little information about the properties
  of the $D_{s}^{\ast}$ mesons is available.
  For example, the quantum number of $J^{P}$, the decay constant
  $f_{D_{s}^{\ast}}$, and the width ${\Gamma}_{D_{s}^{\ast}}$ have
  not yet been determined or confirmed explicitly by experiments.
  It is generally thought that the $J^{P}$ of the $D_{s}^{\ast}$
  mesons is consistent with $1^{-}$ from decay modes \cite{PhysRevLett.75.3232}.
  Some theoretical results on the decay constant $f_{D_{s}^{\ast}}$
  are listed in Table \ref{tab:csstar-decay-constant}.
  It can be seen that the theoretical results are various.
  The recent LQCD results on the decay constant from
  ETM \cite{PhysRevD.96.034524}, HPQCD \cite{PhysRevLett.112.212002}
  and ${\chi}$QCD \cite{cpc.45.023109} groups are in
  reasonable agreement with each other within an error range.
  The latest decay constant $f_{D_{s}^{\ast}}$ $=$ $274{\pm}7$
  MeV from LQCD calculation \cite{cpc.45.023109} will be
  used for an estimation for PLDCV of the $D_{s}^{\ast}$ mesons
  in this paper.
  The experimental upper limit of the decay width is
  ${\Gamma}_{D_{s}^{\ast}}$ $<$ $1.9$ MeV at the 90\,\%
  confidence level set by the CLEO collaboration in 1995
  \cite{PhysRevLett.75.3232}.
  An approximate relation for the decay width,
  ${\Gamma}_{D_{s}^{\ast}}$ ${\simeq}$
  ${\Gamma}(D_{s}^{\ast}{\to}{\gamma}D_{s})$,
  is often used in theoretical calculation.
  The radiative transition process,
  $D_{s}^{\ast}$ ${\to}$ ${\gamma}D_{s}$,
  is a parity conserving decay.
  The parity and angular momentum conservation implies that
  the orbital angular momentum of final states $L$ $=$ $1$.
  There are many theoretical calculation on the decay width
  ${\Gamma}_{D_{s}^{\ast}}$, for example, Refs.
  \cite{PhysRevLett.112.212002,PhysRevD.18.2537,PhysRevD.21.203,
  PhysRevD.31.1081,PhysRevD.37.2564,plb.284.421,plb.334.169,
  PhysRevD.47.1030,plb.316.555,plb.334.175,plb.336.113,PhysRevD.49.299,
  zpc.67.633,PhysRevD.52.6383,mpla.12.3027,npa.658.249,
  npa.671.380,jpg.27.1519,PhysRevD.64.094007,epja.13.363,plb.537.241,
  PhysRevD.68.054024,PhysRevD.72.094004,
  ijmpa.25.2063,epjc.75.243,epja.52.90,ijmpa.31.1650109,
  PhysRevD.94.113011,PhysRevD.100.016019,
  PhysRevD.101.054019,jhep.2020.04.023,2106.13617}.
  The partial decay width for the magnetic dipole transition
  is generally written as \cite{fayyazuddin},
   \begin{equation}
  {\Gamma}(V{\to}P{\gamma})\, =\,
   \frac{4}{3}\,{\alpha}_{\rm em}\,k_{\gamma}^{3}\,{\mu}_{VP}^{2}
   \label{eq:decay-width-vpr},
   \end{equation}
  with the definition of the magnetic dipole moment ${\mu}_{VP}$
  and the momentum of photon $k_{\gamma}$ in the rest frame of the
  vector meson,
   \begin{equation}
  {\mu}_{VP}
   \, =\, {\langle}P{\vert}\hat{\mu}_{z}{\vert}V{\rangle}
   \, =\, {\langle}P{\vert}\sum\limits_{i}
   \frac{Q_{i}}{2\,m_{i}}\,\hat{\sigma}_{z}{\vert}V{\rangle}
   \label{eq:m1-vpr},
   \end{equation}
   \begin{equation}
   k_{\gamma}\, =\, \frac{m_{V}^{2}-m_{P}^{2}}{2\,m_{V}}
   \label{eq:momentum-photon},
   \end{equation}
  where $Q_{i}$ and $m_{i}$ are the electric charge in the unit
  of ${\vert}e{\vert}$ and mass of the constituent quark,
  respectively. With $m_{d}$ ${\approx}$ $336$ MeV,
  $m_{s}$ ${\approx}$ $490$ MeV,
  $m_{c}$ ${\approx}$ $1500$ MeV and the
   \begin{equation}
   {\mu}_{D_{d}^{\ast}D_{d}}
    \, =\, \frac{1}{6} \Big( \frac{2}{m_{c}}-\frac{1}{m_{d}} \Big)
   \label{eq:m1-cd},
   \end{equation}
   \begin{equation}
   {\mu}_{D_{s}^{\ast}D_{s}}
    \, =\, \frac{1}{6} \Big( \frac{2}{m_{c}}-\frac{1}{m_{s}} \Big)
   \label{eq:m1-cs},
   \end{equation}
  one can obtain ${\Gamma}(D_{d}^{\ast}{\to}{\gamma}D_{d})$
  ${\approx}$ $1.8$ keV and
  ${\Gamma}(D_{s}^{\ast}{\to}{\gamma}D_{s})$
  ${\approx}$ $0.36$ keV \cite{fayyazuddin}.
  The theoretical value of partial decay width
  ${\Gamma}(D_{d}^{\ast}{\to}{\gamma}D_{d})$ is roughly
  consistent with the corresponding experimental data
  ${\Gamma}(D_{d}^{\ast}{\to}{\gamma}D_{d})$ $=$
  ${\Gamma}_{D_{d}^{\ast}}{\times}{\cal B}(D_{d}^{\ast}{\to}{\gamma}D_{d})$
  $=$ $1.33{\pm}0.33$ keV within $2\,{\sigma}$ regions \cite{pdg2020}.
  For the moment, we will use ${\Gamma}_{D_{s}^{\ast}}$ $=$
  $0.36$ keV in the calculation to give an estimate of branching
  ratios for the $D_{s}^{{\ast}-}$ ${\to}$ ${\ell}^{-}\bar{\nu}_{\ell}$
  decays.
   \begin{eqnarray}
  {\Gamma}(D_{s}^{{\ast}-}{\to}{\ell}^{-}\bar{\nu}_{\ell})
   &=& 2.4{\pm}0.1\, {\rm meV}, \quad\text{for}\
   {\ell}\, =\, e,\,{\mu}
   \label{eq:width-cs-e}, \\
  {\Gamma}(D_{s}^{{\ast}-}{\to}{\tau}^{-}\bar{\nu}_{\tau})
   &=& 0.28{\pm}0.02\, {\rm meV}
   \label{eq:width-cs-tau}, \\
   {\cal B}(D_{s}^{{\ast}-}{\to}{\ell}^{-}\bar{\nu}_{\ell})
   &=& (6.7{\pm}0.4){\times}10^{-6}, \quad\text{for}\
   {\ell}\, =\, e,\,{\mu}
   \label{eq:br-cs-e}, \\
   {\cal B}(D_{s}^{{\ast}-}{\to}{\tau}^{-}\bar{\nu}_{\tau})
   &=& (7.8{\pm}0.4){\times}10^{-7}
   \label{eq:br-cs-tau}.
   \end{eqnarray}
  If considering the experimental measurement efficiency,
  there are at least more than $10^{7}$ $D_{s}^{\ast}$
  events to experimentally study the $D_{d}^{{\ast}-}$ ${\to}$
  ${\ell}^{-}\bar{\nu}_{\ell}$ decays. And more than $10^{8}$
  $D_{s}^{\ast}$ events might be needed to explore the
  $D_{s}^{{\ast}-}$ ${\to}$ ${\tau}^{-}\bar{\nu}_{\tau}$ decay.

  \begin{figure}[ht]
  \includegraphics[width=0.4\textwidth]{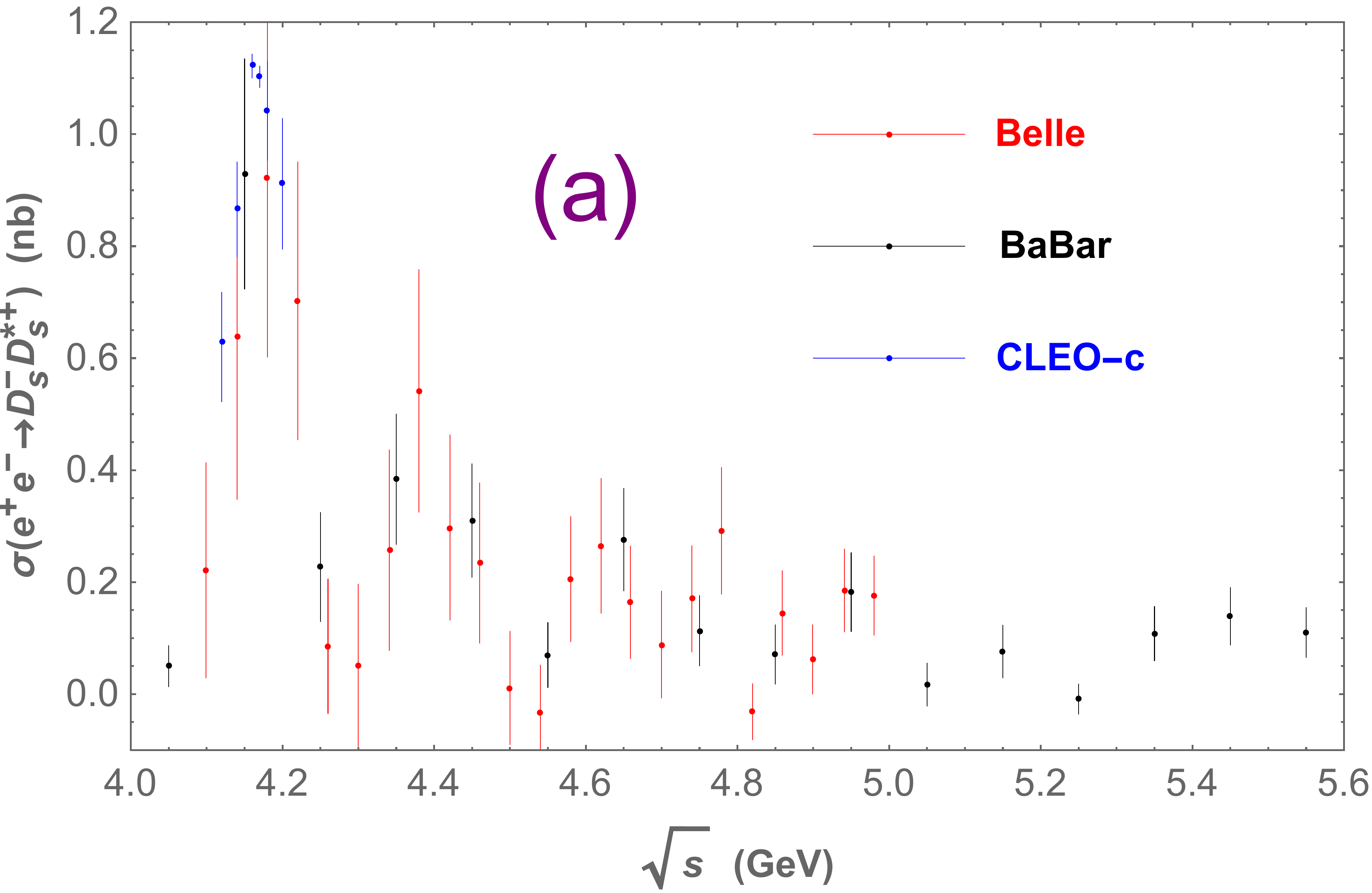}\quad
  \includegraphics[width=0.4\textwidth]{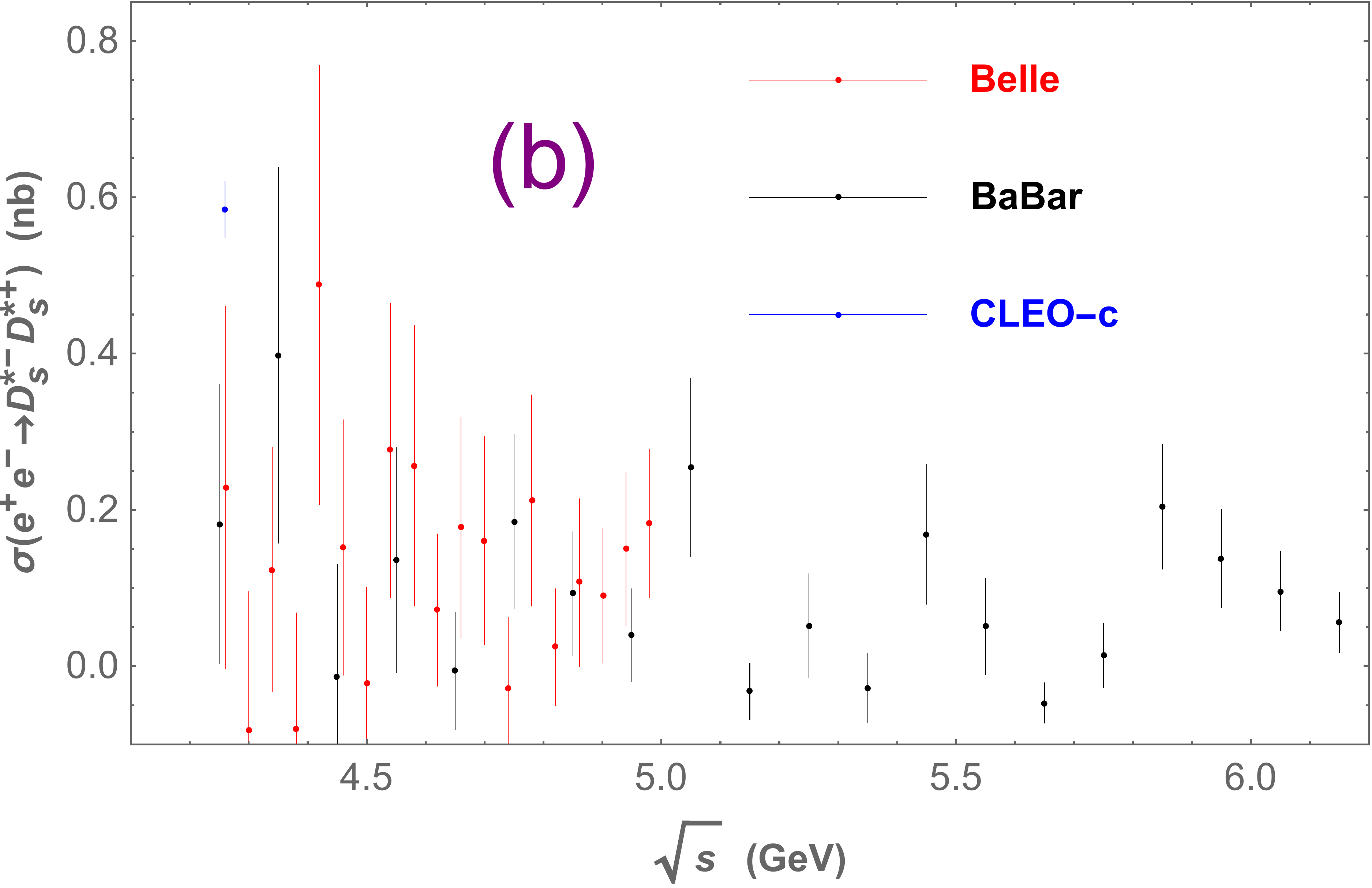}
  \caption{The exclusive cross sections (in the unit of nb)
  as functions of $\sqrt{s}$ (in the unit of GeV) for
  $e^{+}e^{-}$ ${\to}$ $D_{s}^{-}D_{s}^{{\ast}+}$
  in (a) and $e^{+}e^{-}$ ${\to}$ $D_{s}^{{\ast}-}D_{s}^{{\ast}+}$
  in (b).
  The Belle, BaBar and CLEO-c data are from Ref.
  \cite{PhysRevD.83.011101}, Ref. \cite{PhysRevD.82.052004}
  and Ref. \cite{cpc.42.043002}, respectively.}
  \label{fig:cs-dsv}
  \end{figure}

  In the electron-positron collisions, the cross sections of
  $D_{s}^{+}D_{s}^{{\ast}-}$ and $D_{s}^{{\ast}+}D_{s}^{{\ast}-}$
  production have been experimentally studied by the
  Belle \cite{PhysRevD.83.011101}, BaBar \cite{PhysRevD.82.052004}
  and CLEO-c \cite{PhysRevD.80.072001,cpc.42.043002} groups,
  as illustrated in Fig.~\ref{fig:cs-dsv}.
  Assuming the exclusive cross sections near threshold
  ${\sigma}(e^{+}e^{-}{\to}D_{s}^{+}D_{s}^{{\ast}-})$ ${\sim}$
  $1.0\,{\rm nb}$ and
  ${\sigma}(e^{+}e^{-}{\to}D_{s}^{{\ast}+}D_{s}^{{\ast}-})$ ${\sim}$
  $0.2\,{\rm nb}$, there will be about $10^{10}$ $D_{s}^{{\ast}{\pm}}$
  events corresponding to a data sample of $10$ ${\rm ab}^{-1}$ at
  STCF, and about $5{\times}10^{10}$ $D_{s}^{{\ast}{\pm}}$
  events corresponding to a data sample of $50$ ${\rm ab}^{-1}$
  at SuperKEKB.
  In addition, considering the branching ratio
  ${\cal B}(Z{\to}c\bar{c})$ $=$ $(12.03{\pm}0.21)\,\%$ \cite{pdg2020}
  and the fragmentation fraction $f(c{\to}D_{s}^{\ast})$ ${\simeq}$
  $5.5\,\%$ \cite{epjc.76.397}, there will be more than $6{\times}10^{9}$
  (and $6{\times}10^{10}$) $D_{s}^{{\ast}{\pm}}$ events corresponding
  to $10^{12}$ \cite{cepc} (and $10^{13}$ \cite{fcc})
  $Z$ bosons at the future CEPC (and FCC-ee).
  So the $D_{s}^{{\ast}-}$ ${\to}$ ${\ell}^{-}\bar{\nu}_{\ell}$
  decays (with ${\ell}$ $=$ $e$, ${\mu}$ and ${\tau}$) could be
  measured at Belle-II, SCTF, CEPC and FCC-ee experiments.

  In hadron-hadron collisions, the inclusive cross sections for
  the $c\bar{c}$ pair production are
  ${\sigma}(pp{\to}c\bar{c}X)$ ${\simeq}$ $2.4$ mb
  at the center-of-mass energy of $\sqrt{s}$ $=$ $13$ TeV at
  LHCb \cite{JHEP.2016.03.159},
  ${\sigma}_{c\bar{c}}^{\rm tot}$ ${\simeq}$ $8.5$ mb
  and $8.6$ mb at $\sqrt{s}$ $=$ $7$ TeV at
  ALICE \cite{JHEP.2012.07.191} and
  ATLAS \cite{npb.907.717}, respectively.
  With the fragmentation fraction $f(c{\to}D_{s}^{\ast})$ ${\simeq}$
  $5.5\,\%$ \cite{epjc.76.397}, there will be about $4{\times}10^{13}$
  $D_{s}^{{\ast}{\pm}}$ events corresponding a data sample of
  $300\,{\rm fb}^{-1}$ at LHCb, and more $D_{s}^{{\ast}{\pm}}$
  events available at ALICE and ATLAS.
  So the $D_{s}^{{\ast}-}$ ${\to}$ $e^{-}\bar{\nu}_{e}$,
  ${\mu}^{-}\bar{\nu}_{\mu}$, ${\tau}^{-}\bar{\nu}_{\tau}$
  decays could be measured precisely at LHCb, ALICE and
  ATLAS experiments.

  \section{$B_{u}^{{\ast}-}$ ${\to}$ ${\ell}^{-}\bar{\nu}_{\ell}$ decays}
  \label{sec-bu}
  The experimental information about the $B_{u}^{\ast}$ mesons are
  very scarce.
  The already known information about the $B_{u}^{\ast}$ mesons
  are their quark composition $b\bar{u}$ with the quark model assignment,
  the isospin $I$ $=$ $1/2$, the spin-parity quantum number $J^{P}$
  $=$ $1^{-}$ and the mass $m_{B_{u}^{\ast}}$ $=$ $5324.70(21)$
  MeV \cite{pdg2020}.
  Due to the mass difference $m_{B_{u}^{\ast}}$ $-$ $m_{B_{u}}$ $=$
  $45$ MeV $<$ $m_{\pi}$, the electromagnetic radiative transition
  $B_{u}^{\ast}$ ${\to}$ $B_{u}{\gamma}$ certainly will be the
  important and dominant decay mode.
  The photon in the $B_{u}^{\ast}$ ${\to}$ $B_{u}{\gamma}$ decay
  is very soft, with the momentum $k_{\gamma}$ ${\sim}$ $45$ MeV
  in the center-of-mass of the $B_{u}^{\ast}$ mesons.
  No signal event of the $B_{u}^{\ast}$ ${\to}$ $B_{u}{\gamma}$
  decay has yet been found.
  The $B_{u}^{{\ast}-}$ ${\to}$ ${\ell}^{-}\bar{\nu}_{\ell}$ decays
  offer a complementary decay modes of the $B_{u}^{\ast}$ meson.
  It can be seen from Eq.(\ref{eq:decay-width--vector}) that
  the information about ${\vert}V_{ub}{\vert}\,f_{B_{u}^{\ast}}$
  could be obtained, however, the partial width for the purely
  leptonic decays $B_{u}^{{\ast}-}$ ${\to}$ ${\ell}^{-}\bar{\nu}_{\ell}$
  are highly suppressed by the CKM element of
  ${\vert}V_{ub}{\vert}^{2}$ ${\sim}$
  ${\cal O}({\lambda}^{6})$.

  The precise determinations of the CKM element $V_{ub}$ $=$
  ${\vert}V_{ub}{\vert}\,e^{-i\,{\gamma}}$ are very
  central and important to verify the CKM picture of SM,
  where ${\gamma}$ is the angle of the unitarity triangle
  of $V_{ud}V_{ub}^{\ast}$ $+$ $V_{cd}V_{cb}^{\ast}$ $+$
  $V_{td}V_{tb}^{\ast}$ $=$ $0$.
  The experimental determination of ${\vert}V_{ub}{\vert}$
  from the inclusive $B$ ${\to}$ $X_{u}{\ell}\bar{\nu}_{\ell}$
  decay is complicated mainly by the large backgrounds from
  the CKM-favored $B$ ${\to}$ $X_{c}{\ell}\bar{\nu}_{\ell}$ decay.
  The experimental extraction of ${\vert}V_{ub}{\vert}$ from
  exclusive $B$ ${\to}$ ${\pi}{\ell}\bar{\nu}_{\ell}$ decay is
  subject to the form factors calculated with the lattice QCD
  or QCD sum rules.
  The latest values obtained from inclusive and exclusive
  determinations are \cite{pdg2020}
   \begin{equation}
  {\vert}V_{ub}{\vert}\,{\times}10^{3}
   \, =\, 4.25{\pm}0.12_{\rm\, exp}
    {}^{+0.15}_{-0.14}{}_{\rm\, theo}
    {\pm}0.23_{\, {\Delta}BF}
   \, =\, 4.25^{+0.30}_{-0.29}
   \quad \text{(inclusive)}
   \label{eq:vub-inclusive},
   \end{equation}
   \begin{equation}
  {\vert}V_{ub}{\vert}\,{\times}10^{3}
   \, =\, 3.70{\pm}0.10_{\rm\, exp}{\pm}0.12_{\rm\, theo}
   \, =\, 3.70{\pm}0.16
   \quad \text{(exclusive)}
   \label{eq:vub-exclusive}.
   \end{equation}
  It is clearly seen that
  (1)
  the difference between inclusive and exclusive determinations
  of ${\vert}V_{ub}{\vert}$ is obvious.
  (2)
  The best determinations of ${\vert}V_{ub}{\vert}$ are from
  exclusive semileptonic decays, with a precision of about
  $4\,\%$. The experimental errors for the exclusive semileptonic
  decays are expected to decrease from the current $2.7\,\%$ to
  $1.2\,\%$ based on a dataset of $50$ ${\rm ab}^{-1}$  at Belle-II
  experiments \cite{PTEP.2019.123C01}.
  (3) The theoretical uncertainties are larger than
  experimental ones.
  Moreover, ${\vert}V_{ub}{\vert}$ can also be experimentally
  determined from the leptonic decay $B_{u}$ ${\to}$
  ${\tau}\bar{\nu}_{\tau}$ and the semileptonic hyperon decay
  ${\Lambda}_{b}$ ${\to}$ $p{\mu}\bar{\nu}_{\mu}$.
  The constraint from the global fit gives ${\vert}V_{ub}{\vert}$
  $=$ $(3.61^{+0.11}_{-0.09}){\times}10^{-3}$ \cite{pdg2020},
  which will be used in our calculation.

   \begin{table}[ht]
   \caption{The theoretical values of decay constant $f_{B_{u}^{\ast}}$
   (in the unit of MeV), where the legends are the same as those in
   Table \ref{tab:cdstar-decay-constant}.}
   \label{tab:bustar-decay-constant}
   \begin{ruledtabular}
   \begin{tabular}{lccccc}
  NRQM & $280$ \cite{plb.635.93}
       & $151^{+15}_{-13}$ \cite{PhysRevD.71.113006}
       & $234.7$ \cite{epjp.132.80} \footnotemark[1]
       & $225.1$ \cite{epjp.132.80} \footnotemark[2] \\
  NRQM & $196$ \cite{ahep.2018.7032041} \footnotemark[1]
       & $182$ \cite{ahep.2018.7032041} \footnotemark[2]
       & $213$ \cite{mpla.17.803}
       & $242.37$ \cite{epl.116.31004} \footnotemark[1]
       & $232.47$ \cite{epl.116.31004} \footnotemark[2] \\
  RQM  & $219$ \cite{plb.635.93}
       & $195$ \cite{mpla.17.803}
       & $252{\pm}10$ \cite{PhysRevD.55.6944} \footnotemark[3]
       & $193{\pm}8$ \cite{PhysRevD.55.6944} \footnotemark[4] \\
  LFQM & $188$ \cite{PhysRevC.92.055203} \footnotemark[5]
       & $198.7^{+~4.9}_{-11.3}$ \cite{cpc.42.073102} \footnotemark[6]
       & $193.1^{+4.3}_{-4.6}$ \cite{cpc.42.073102} \footnotemark[7]
       & $196^{+28}_{-27}$ \cite{jpg.39.025005}
       & $172^{+24}_{-23}{\pm}6$ \cite{PhysRevD.100.014026} \\
  LFQM & $204$ \cite{PhysRevD.75.073016} \footnotemark[8]
       & $193$ \cite{PhysRevD.75.073016} \footnotemark[9]
       & $225{\pm}38$ \cite{PhysRevD.81.114024} \footnotemark[10]
       & $249^{+44}_{-42}$ \cite{PhysRevD.81.114024} \footnotemark[11]
       & $205{\pm}5$ \cite{PhysRevD.98.114018} \\
  LFQM & $201.9^{+43.2}_{-41.4}$ \cite{epjc.76.313} \footnotemark[10]
       & $220.2^{+49.1}_{-46.2}$ \cite{epjc.76.313} \footnotemark[11]
       & $227$ \cite{epjp.133.134} \footnotemark[12]
       & $211$ \cite{epjp.133.134} \footnotemark[13]
       & $228$ \cite{epjp.133.134} \footnotemark[14] \\
  LQCD  & $196{\pm}24^{+39}_{-~2}$ \cite{PhysRevD.60.074501}
        & $190{\pm}28$ \cite{npb.619.507} \footnotemark[15]
        & $185.9{\pm}7.2$ \cite{PhysRevD.96.034524}
        & $175{\pm}6$ \cite{PhysRevD.91.114509} \\
  SR    & $210^{+10}_{-12}$ \cite{PhysRevD.88.014015}
        & $213{\pm}18$ \cite{epjc.75.427}
        & $209{\pm}8$ \cite{ijmpa.30.1550116}
        & $209^{+23}_{-22}$ \cite{2106.13617}
        & $181.8{\pm}13.7$ \cite{PhysRevD.91.116009} \\
  other & $138$ \cite{epjp.133.134} \footnotemark[16]
        & $200{\pm}10$ \cite{PhysRevD.75.116001}
        & $238{\pm}18$ \cite{plb.633.492}
        & $164$ \cite{PhysRevD.58.014007}
  \end{tabular}
  \end{ruledtabular}
  \footnotetext[1]{Without QCD radiative corrections.}
  \footnotetext[2]{With QCD radiative corrections.}
  \footnotetext[3]{With $f_{B_{u}^{\ast}}$ $=$ $f_{B_{d}^{\ast}}$, and
                   the constituent quark masses for the
                   light quarks $s$ and $d$.}
  \footnotetext[4]{With $f_{B_{u}^{\ast}}$ $=$ $f_{B_{d}^{\ast}}$, and
                   the current quark masses for the light quarks
                   $s$ and $d$.}
  \footnotetext[5]{With Coulomb plus linear potential model.}
  \footnotetext[6]{With a dilation parameter ${\kappa}$ $=$ $0.54$ GeV.}
  \footnotetext[7]{With a dilation parameter ${\kappa}$ $=$ $0.68$ GeV.}
  \footnotetext[8]{With Coulomb plus linear potential model.}
  \footnotetext[9]{With Coulomb plus harmonic oscillator potential model.}
  \footnotetext[10]{With the Gaussian type wave functions.}
  \footnotetext[11]{With the power-law type wave functions.}
  \footnotetext[12]{With Martin potential model \cite{plb.93.338}.}
  \footnotetext[13]{With Cornell potential model \cite{zpc.33.135}.}
  \footnotetext[14]{With logarithmic potential model \cite{plb.71.153}.}
  \footnotetext[15]{With $m_{B^{\ast}}/f_{B^{\ast}}$ $=$ $28{\pm}1^{+3}_{-4}$
                    \cite{npb.619.507} and $m_{B^{\ast}}$ $=$ $5324.70(21)$ MeV
                    \cite{pdg2020}.}
  \footnotetext[16]{With harmonic plus Yukawa potential model \cite{epjp.132.80}.}
  \end{table}

  Some theoretical results on the decay constant $f_{B_{u}^{\ast}}$
  are collected in Table \ref{tab:bustar-decay-constant}.
  The recent result $f_{B_{u}^{\ast}}$ $=$ $185.9{\pm}7.2$ MeV from
  LQCD calculation \cite{PhysRevD.96.034524} will be used in this
  paper.
  There are many theoretical calculation on the decay width
  ${\Gamma}_{B_{u}^{\ast}}$, for example, Refs.
  \cite{PhysRevD.47.1030,plb.316.555,plb.334.175,PhysRevD.49.299,
  zpc.67.633,mpla.12.3027,npa.658.249,npa.671.380,
  jpg.27.1519,PhysRevD.64.094007,epja.13.363,plb.537.241,
  PhysRevD.68.054024,ijmpa.25.2063,epja.52.90,PhysRevD.94.113011,
  PhysRevD.100.016019,jhep.2020.04.023,2106.13617}.
  With the formula of Eq.(\ref{eq:decay-width-vpr}), the quark
  mass $m_{u}$ $=$ $336$ MeV \cite{fayyazuddin} and
  $m_{b}$ $=$ $4.78$ GeV \cite{pdg2020}, and the magnetic dipole
  momentum
   \begin{equation}
   {\mu}_{B_{u}^{\ast}B_{u}}
    \, =\, \frac{1}{6} \Big( \frac{2}{m_{u}}-\frac{1}{m_{b}} \Big)
   \label{eq:m1-bu},
   \end{equation}
  one can obtain ${\Gamma}_{B_{u}^{\ast}}$ ${\simeq}$
  ${\Gamma}(B_{u}^{\ast}{\to}{\gamma}B_{u})$ ${\simeq}$
  $820$ eV.
  The partial decay width and branching ratios for the
  $B_{u}^{{\ast}-}$ ${\to}$ ${\ell}^{-}\bar{\nu}_{\ell}$ decays are
   \begin{eqnarray}
  {\Gamma}(B_{u}^{{\ast}-}{\to}{\ell}^{-}\bar{\nu}_{\ell})
   &=& 0.25^{+0.04}_{-0.03}\, {\mu}\,{\rm eV}, \quad\text{for}\
   {\ell}\, =\, e,\,{\mu}
   \label{eq:width-bu-e}, \\
  {\Gamma}(B_{u}^{{\ast}-}{\to}{\tau}^{-}\bar{\nu}_{\tau})
   &=& 0.20^{+0.03}_{-0.02}\, {\mu}\,{\rm eV}
   \label{eq:width-bu-tau}, \\
   {\cal B}(B_{u}^{{\ast}-}{\to}{\ell}^{-}\bar{\nu}_{\ell})
   &=& (3.0{\pm}0.4){\times}10^{-10}, \quad\text{for}\
   {\ell}\, =\, e,\,{\mu}
   \label{eq:br-bu-e}, \\
   {\cal B}(B_{u}^{{\ast}-}{\to}{\tau}^{-}\bar{\nu}_{\tau})
   &=& (2.5{\pm}0.4){\times}10^{-10}
   \label{eq:br-bu-tau}.
   \end{eqnarray}
  It is expected that there should be at least more than $10^{11}$
  $B_{u}^{\ast}$ events available for experimental study of the
  $B_{u}^{{\ast}-}$ ${\to}$ ${\ell}^{-}\bar{\nu}_{\ell}$ decays.

  The experimental study has shown that the exclusive cross sections
  for the final states of $B\overline{B}^{\ast}$,
  $B^{\ast}\overline{B}^{\ast}$ and $B\overline{B}^{\ast}{\pi}$
  will have a large share of the total $b\bar{b}$ cross sections
  above the open bottom threshold, for example \cite{pdg2020},
   \begin{eqnarray}
   & &
  {\cal B}(Z_{b}(10610){\to}B^{+}\overline{B}^{{\ast}0}+B^{{\ast}+}\overline{B}^{0})
   \, =\, 85.6^{+2.1}_{-2.9} \,\%
   \label{eq:upsilion-10610-b-bv}, \\
   & &
  {\cal B}(Z_{b}(10650){\to}B^{{\ast}+}\overline{B}^{{\ast}0})
   \, =\, 74^{+4}_{-6} \,\%
   \label{eq:upsilion-10650-bv-bv}, \\
   & &
  {\cal B}({\Upsilon}(5S){\to}B\overline{B}^{\ast}+c.c.)
   \, =\, 13.7{\pm}1.6 \,\%
   \label{eq:upsilion-5s-b-bv}, \\
   & &
  {\cal B}({\Upsilon}(5S){\to}B^{\ast}\overline{B}^{\ast})
   \, =\,  38.1{\pm}3.4 \,\%
   \label{eq:upsilion-5s-bv-bv}, \\
   & &
  {\cal B}({\Upsilon}(5S){\to}B\overline{B}^{\ast}{\pi}
    +B^{\ast}\overline{B}{\pi})
   \, =\,  7.3{\pm}2.3\,\%
   \label{eq:upsilion-5s-b-bv-pi}, \\
   & &
  {\cal B}({\Upsilon}(5S){\to}B^{\ast}\overline{B}^{\ast}{\pi})
   \, =\, 1.0{\pm}1.4\,\%
   \label{eq:upsilion-5s-bv-bv-pi}.
   \end{eqnarray}
  There are about $36{\times}10^{6}$ ${\Upsilon}(5S)$ events
  corresponding to the dataset of $121$ ${\rm fb}^{-1}$
  at Belle experiments at the disposal \cite{PTEP.2019.123C01}.
  About $1.5{\times}10^{10}$ ${\Upsilon}(5S)$ events
  with a dataset $50$ ${\rm ab}^{-1}$ at Belle-II are an
  outside estimate. Assuming the inclusive branching ratio
  ${\cal B}({\Upsilon}(5S){\to}B_{u}^{\ast}X)$ ${\simeq}$
  $30\,\%$, there will be some $4.5{\times}10^{9}$ $B_{u}^{\ast}$
  events at most at Belle-II. And it is more important that the vast
  majority of the data will be taken at ${\Upsilon}(4S)$ resonance
  rather than ${\Upsilon}(5S)$ mesons at Belle-II experiments,
  and ${\Upsilon}(4S)$ lies below the $B\overline{B}^{\ast}$
  threshold.
  So the probability of direct observation of the $B_{u}^{{\ast}-}$
  ${\to}$ ${\ell}^{-}\bar{\nu}_{\ell}$ decays at Belle-II experiments
  should be very tiny.
  Considering about $10^{13}$ $Z$ bosons at FCC-ee \cite{fcc}
  and branching ratio ${\cal B}(Z{\to}b\bar{b})$ $=$
  $12.03{\pm}0.21\,\%$ \cite{pdg2020}, and assuming the
  fragmentation fraction $f(b{\to}B_{u}^{\ast})$ ${\sim}$
  $20\,\%$ \cite{plb.576.29}, there will be more than
  $4{\times}10^{11}$ $B_{u}^{\ast}$
  events to search for the $B_{u}^{{\ast}-}$ ${\to}$
  ${\ell}^{-}\bar{\nu}_{\ell}$ decays.
  The $b$-quark production cross sections at the center-of-mass
  energy $\sqrt{s}$ $=$ 13 TeV is about ${\sigma}(pp{\to}b\bar{b}X)$
  ${\simeq}$ $495$ ${\mu}$b at LHCb \cite{JHEP.2015.10.172}.
  There will be more than $5{\times}10^{13}$ $B_{u}^{\ast}$ events
  with a dateset of $300\,{\rm fb}^{-1}$ at LHCb and fragmentation
  fraction $f(b{\to}B_{u}^{\ast})$ ${\sim}$ $20\,\%$.
  Hence, the $B_{u}^{{\ast}-}$ ${\to}$ $e^{-}\bar{\nu}_{e}$,
  ${\mu}^{-}\bar{\nu}_{\mu}$, ${\tau}^{-}\bar{\nu}_{\tau}$
  decays could be investigated at FCC-ee and LHCb experiments
  in the future.

  \section{$B_{c}^{{\ast}-}$ ${\to}$ ${\ell}^{-}\bar{\nu}_{\ell}$ decays}
  \label{sec-bu}
  According to the conventional quark-model assignments, the
  $B_{c}^{\ast}$ mesons consist of two heavy quarks with different
  flavor numbers $B$ $=$ $C$ $=$ $-Q$ $=$ ${\pm}1$.
  Up to today, the experimental information of the $B_{c}^{\ast}$
  meson is still very limited.
  For example, the potential candidate of the $B_{c}^{\ast}$ meson
  has not yet been determined.
  It is generally believed that the mass of the $B_{c}^{\ast}$ meson
  should be in the region between $m_{B_{c}}$ $=$
  $6274.47{\pm}0.27{\pm}0.17$ MeV recently measured by
  LHCb \cite{JHEP.2020.07.123} and $m_{B_{c}(2S)}$ $=$
  $6872.1{\pm}1.3{\pm}0.1{\pm}0.8$ MeV obtained by LHCb
  \cite{PhysRevLett.122.232001} (or $6871.0{\pm}1.2{\pm}0.8{\pm}0.8$
  MeV given by CMS \cite{PhysRevLett.122.132001}),
  where the $B_{c}$, $B_{c}^{\ast}$ and $B_{c}(2S)$ particles
  correspond to the sibling isoscalar states with quantum numbers of
  $n^{2S+1}L_{J}$ $=$ $1^{1}S_{0}$, $1^{3}S_{1}$ and $2^{1}S_{0}$,
  respectively.
  So the branching ratios for the strong decays $B_{c}^{\ast}$
  ${\to}$ $BD$ are zero, because $B_{c}^{\ast}$ meson is below
  the $BD$ pair threshold.
  The experimental particle physicists are earnestly looking for and
  identifying the $B_{c}^{\ast}$ meson, a long-expected charming beauty.
  For the moment, almost all of the information available about the
  properties of $B_{c}^{\ast}$ meson (such as the mass, decay
  constant, lifetime, decay modes and so on) come from theoretical
  estimates.
  There are too many estimations on the $B_{c}^{\ast}$ meson mass
  with various theoretical models, for example, in Refs.
  \cite{zpc.3.165,PhysRevD.21.3180,PhysRevD.23.2724,PhysRevD.24.132,
  zpc.12.63,PhysRevD.32.189,ijmpa.6.2309,PhysRevD.44.212,
  PhysRevD.46.1165,PhysRevD.49.5845,zpc.64.57,Phys.Usp.38.1,
  PhysRevD.51.1248,PhysRevD.51.3613,PhysRevD.52.5229,
  PhysRevD.53.312,plb.382.131,epjc.4.107,PhysRevD.59.094001,
  PhysRevD.60.074006,PhysRevD.62.114024,PhysRevD.70.054017,
  ijmpa.19.1771,epjc.37.323,PhysRevD.71.034006,
  Pramana.66.953,plb.651.171,PhysRevC.78.055202,jpg.36.035003,
  npa.848.299,PhysRevD.81.076005,
  PhysRevLett.104.022001,PhysRevD.81.071502,epjc.71.1825,
  pan.74.631,PhysRevD.86.094510,epja.49.131,epja.50.154,
  ijmpa.32.1750021,PhysRevLett.121.202002,epjc.78.592,
  PhysRevD.99.054025,PhysRevD.99.096020,
  PhysRevD.100.096002,PhysRevD.100.114032,epja.55.82,
  npb.947.114726,epjc.80.223,
  plb.807.135522,PhysRevD.101.056002,PhysRevD.102.034002,
  pan.83.634,plb.816.136277,fewbody.62.39,epjc.81.327}.
  The recent result from lattice QCD calculation,
  $m_{B_{c}^{\ast}}$ $=$ $6331{\pm}7$ MeV \cite{PhysRevLett.121.202002},
  which are basically consistent with other estimations,
  will be used in this paper.
  Clearly, it is foreseeable that the isospin violating decay
  $B_{c}^{\ast}$ ${\to}$ $B_{c}{\pi}$ is explicitly forbidden
  by the law of energy conservation, because of $m_{B_{c}^{\ast}}$
  $-$ $m_{B_{c}}$ ${\simeq}$ $57$ MeV $<$ $m_{\pi}$.
  Hence, the electromagnetic radiative transition $B_{c}^{\ast}$
  ${\to}$ $B_{c}{\gamma}$ should be the dominant decay mode.
  In addition, the photon in the magnetic dipole transition
  $B_{c}^{\ast}$ ${\to}$ $B_{c}{\gamma}$ is very soft in the
  rest frame of the $B_{c}^{\ast}$ meson.
  This might be one main reason why the unambiguously experimental
  identification of the $B_{c}^{\ast}$ meson is very challenging.
  As an important complementary decay modes, the $B_{c}^{\ast}$
  meson has very rich weak decay channels, which could be
  approximately classified into three classes:
  (1) the valence $b$ quark weak decay accompanied by the spectator $c$ quark,
  (2) the valence $c$ quark weak decay accompanied by the spectator $b$ quark,
  and (3) the $b$ and $c$ quarks annihilation into a virtual $W$ boson.
  The purely leptonic decays $B_{c}^{{\ast}-}$ ${\to}$
  ${\ell}^{-}\bar{\nu}_{\ell}$ belong to the third case,
  which are favored by the CKM element ${\vert}V_{cb}{\vert}$.
  And the information of ${\vert}V_{cb}{\vert}\,f_{B_{c}^{\ast}}$
  could be obtained from the $B_{c}^{{\ast}-}$ ${\to}$
  ${\ell}^{-}\bar{\nu}_{\ell}$ decays.

   \begin{table}[ht]
   \caption{The theoretical values of decay constant $f_{B_{c}^{\ast}}$
   (in the unit of MeV), where the legends are the same as those in
   Table \ref{tab:cdstar-decay-constant}.}
   \label{tab:bcstar-decay-constant}
   \begin{ruledtabular}
   \begin{tabular}{lcccccc}
  NRQM & $562$ \cite{mpla.17.803} %
       & $434.64$ \cite{epjc.78.592} %
       & $544.3$ \cite{pan.83.634} \\ %
  RQM  & $503$ \cite{mpla.17.803}  %
       & $510{\pm}80$ \cite{PhysRevD.51.3613} \footnotemark[1] %
       & $456{\pm}70$ \cite{PhysRevD.51.3613} \footnotemark[2] %
       & $460{\pm}60$ \cite{PhysRevD.51.3613} \footnotemark[3] \\ %
  LFQM & $391^{+4}_{-5}$ \cite{PhysRevC.92.055203} %
       & $440^{+51}_{-52}$ \cite{jpg.39.025005} %
       & $387$ \cite{PhysRevD.81.114024} \footnotemark[4] %
       & $423$ \cite{PhysRevD.81.114024} \footnotemark[5]
       & $465{\pm}7$ \cite{PhysRevD.98.114018} \\
  LFQM & $473.4{\pm}18.2$ \cite{epjc.76.313} \footnotemark[4] %
       & $487.6{\pm}19.2$ \cite{epjc.76.313} \footnotemark[5] %
       & $398$ \cite{PhysRevD.80.054016} \footnotemark[6] %
       & $551$ \cite{PhysRevD.80.054016} \footnotemark[7] %
       & $474{\pm}42$ \cite{PhysRevD.98.114038} \\
  LQCD & $422{\pm}13$ \cite{PhysRevD.91.114509} %
       & $387{\pm}12$ \cite{lattice2018.273} \\
  SR    & $384{\pm}32$ \cite{epja.49.131} \footnotemark[8] %
        & $415{\pm}31$ \cite{epja.49.131} \footnotemark[9] %
        & $300{\pm}30$ \cite{npb.947.114726} %
        & $442{\pm}44$ \cite{plb.807.135522} \footnotemark[10] %
        & $387{\pm}15$ \cite{plb.807.135522} \footnotemark[11] %
        \\
  other & $453{\pm}20$ \cite{PhysRevD.75.116001} %
        & $418{\pm}24$ \cite{plb.633.492} %
        & $471$ \cite{PhysRevD.99.054025} %
  \end{tabular}
  \end{ruledtabular}
  \footnotetext[1]{With Martin potential model \cite{plb.93.338}.}
  \footnotetext[2]{With Coulomb plus linear potential model.}
  \footnotetext[3]{Obtained from the scaling relation.}
  \footnotetext[4]{With the Gaussian type wave functions.}
  \footnotetext[5]{With the power-law type wave functions.}
  \footnotetext[6]{With Coulomb plus linear potential model.}
  \footnotetext[7]{With Coulomb plus harmonic oscillator potential model.}
  \footnotetext[8]{With the current quark mass.}
  \footnotetext[9]{With the pole quark mass.}
  \footnotetext[10]{With inputs from the inverse Laplace-type model.}
  \footnotetext[11]{With inputs from Heavy Quark Symmetry.}
  \end{table}

  The current values of the CKM element ${\vert}V_{cb}{\vert}$
  come mainly from inclusive and exclusive semileptonic
  decays of $B$ meson to charm \cite{pdg2020}.
  The average values obtained from inclusive $b$ ${\to}$
  $c{\ell}\bar{\nu}_{\ell}$ decays and exclusieve
  $B$ ${\to}$ $D^{(\ast)}{\ell}\bar{\nu}_{\ell}$ decays are
  ${\vert}V_{cd}{\vert}{\times}10^{3}$ $=$ $42.2(8)$ and
  $39.5(9)$, respectively \cite{pdg2020}.
  The lepton flavor non-universality in the ratio $R(D^{({\ast})})$
  complicate the determination of ${\vert}V_{cb}{\vert}$.
  In addition, ${\vert}V_{cb}{\vert}$ can also be obtained
  from the PLDCM $B_{c}^{-}$ ${\to}$ ${\ell}\bar{\nu}_{\ell}$
  decays, although none of the measurements has reached a
  competitive level of precision due to either the serious
  helicity suppression for $B_{c}^{-}$ ${\to}$ $e\bar{\nu}_{e}$,
  ${\mu}\bar{\nu}_{\mu}$ decays or other additional neutrinos
  from ${\tau}$ decay for $B_{c}^{-}$ ${\to}$
  ${\tau}\bar{\nu}_{\tau}$ decay.
  The global SM fit value is ${\vert}V_{cb}{\vert}$ $=$
  $40.53^{+0.83}_{-0.61}{\times}10^{-3}$ \cite{pdg2020},
  which will be used in this paper.

  Both valence quarks of the $B_{c}^{({\ast})}$ mesons are
  regarded as heavy quarks.
  Their Compton wave lengths ${\sim}$ $1/m_{b,c}$ are much
  shorter than a typical hadron size.
  The spin-flavor symmetry in the heavy quark limit would
  lead to an approximation between decay constants
  $f_{B_{c}}$ ${\approx}$ $f_{B_{c}^{\ast}}$.
  Some theoretical results on the decay constant $f_{B_{c}^{\ast}}$
  are collected in Table \ref{tab:bcstar-decay-constant}.
  The recent lattice QCD calculation $f_{B_{c}^{\ast}}$ $=$
  $387{\pm}12$ MeV \cite{lattice2018.273} will be used in this paper.
  As it is well known that the magnetic momentum of both $b$ and
  $c$ quarks are inversely proportional to their mass.
  The magnetic dipole momentum
   \begin{equation}
   {\mu}_{B_{c}^{\ast}B_{c}}
    \, =\, \frac{1}{6} \Big( \frac{2}{m_{c}}-\frac{1}{m_{b}} \Big)
   \label{eq:m1-bc},
   \end{equation}
  should be very small.
  With the quark mass $m_{c}$ $=$ $1.5$ GeV and $m_{b}$ $=$
  $4.78$ GeV, one can obtain ${\Gamma}_{B_{c}^{\ast}}$ ${\simeq}$
  ${\Gamma}(B_{c}^{\ast}{\to}{\gamma}B_{c})$ ${\simeq}$
  $60$ eV using the formula of Eq.(\ref{eq:decay-width-vpr}).
  The partial decay width and branching ratios for the
  $B_{c}^{{\ast}-}$ ${\to}$ ${\ell}^{-}\bar{\nu}_{\ell}$ decays
  are estimated to be,
   \begin{eqnarray}
  {\Gamma}(B_{c}^{{\ast}-}{\to}{\ell}^{-}\bar{\nu}_{\ell})
   &=& 225^{+25}_{-21}\, {\mu}\,{\rm eV}, \quad\text{for}\
   {\ell}\, =\, e,\,{\mu}
   \label{eq:width-bc-e}, \\
  {\Gamma}(B_{c}^{{\ast}-}{\to}{\tau}^{-}\bar{\nu}_{\tau})
   &=& 198^{+22}_{-18}\, {\mu}\,{\rm eV}
   \label{eq:width-bc-tau}, \\
   {\cal B}(B_{c}^{{\ast}-}{\to}{\ell}^{-}\bar{\nu}_{\ell})
   &=& (3.8^{+0.4}_{-0.3}){\times}10^{-6}, \quad\text{for}\
   {\ell}\, =\, e,\,{\mu}
   \label{eq:br-bc-e}, \\
   {\cal B}(B_{c}^{{\ast}-}{\to}{\tau}^{-}\bar{\nu}_{\tau})
   &=& (3.3^{+0.4}_{-0.3}){\times}10^{-6}
   \label{eq:br-bc-tau}.
   \end{eqnarray}
  To experimentally investigate the $B_{c}^{{\ast}-}$ ${\to}$
  ${\ell}^{-}\bar{\nu}_{\ell}$ decays, there should be at
  least more than $10^{7}$ $B_{c}^{\ast}$ events available.

  More than $10^{12}$ $Z$ bosons are expected at the future
  $e^{+}e^{-}$ colliders of CEPC \cite{cepc} and FCC-ee \cite{fcc}.
  Considering the branching ratio ${\cal B}(Z{\to}b\bar{b})$ $=$
  $12.03{\pm}0.21\,\%$ \cite{pdg2020} and fragmentation
  fraction $f(b{\to}B_{c}^{\ast})$ ${\sim}$ $6{\times}10^{-4}$
  \cite{npa.953.21,cpc.43.083101,PhysRevD.100.034004},
  there will be more than $10^{8}$ $B_{c}^{\ast}$ events to
  search for the $B_{c}^{{\ast}-}$ ${\to}$ $e^{-}\bar{\nu}_{e}$,
  ${\mu}^{-}\bar{\nu}_{\mu}$, ${\tau}^{-}\bar{\nu}_{\tau}$
  decays.
  In addition, the $B_{c}^{\ast}$ production cross sections
  at LHC are estimated to be about $100$ nb for $pp$ collisions
  at $\sqrt{s}$ $=$ $13$ TeV, about $8$ mb for $p$-Pb collisions
  at $\sqrt{s}$ $=$ $8.16$ TeV and some $920$ mb for Pb-Pb
  collisions at $\sqrt{s}$ $=$ $5.02$ TeV, respectively
  \cite{PhysRevD.97.114022}.
  There will be more than $3{\times}10^{10}$ $B_{c}^{\ast}$ events
  corresponding to a dataset of $300\,{\rm fb}^{-1}$ at LHCb
  for $pp$ collisions.
  Hence, the $B_{c}^{{\ast}-}$ ${\to}$ $e^{-}\bar{\nu}_{e}$,
  ${\mu}^{-}\bar{\nu}_{\mu}$, ${\tau}^{-}\bar{\nu}_{\tau}$
  decays are expected to be carefully measured at LHCb
  experiments in the future.

  \section{summary}
  \label{sec-sum}
  The mass of the charged vector mesons are generally larger
  than that of the corresponding ground pseudoscalar mesons.
  The vector mesons decay mainly through the strong or/and
  electromagnetic interactions.
  These facts will inevitably result in that the branching ratios
  of the vector meson weak decays are often very tiny.
  Inspired by the potential prospects of existing and coming
  high-luminosity experiments, more and more experimental data
  will be accumulated, and higher measurement precision level will
  be reached.
  The probabilities of experimental investigation on the purely
  leptonic decays of charged vector mesons are discussed in
  this paper.
  We found that
  (1) for both ${\rho}^{\pm}$ and $K^{{\ast}{\pm}}$ mesons, their
  widths are large due to the dominance of strong decay.
  Their PLDCV branching ratios are estimated at the order
  of ${\cal O}(10^{-13})$.
  Although extremely complicated and difficult, the PLDCV
  decays ${\rho}^{\pm}$, $K^{{\ast}{\pm}}$ ${\to}$
  $e^{-}\bar{\nu}_{e}$, ${\mu}^{-}\bar{\nu}_{\mu}$ might be
  measurable due to the huge data of the ${\rho}^{\pm}$
  and $K^{{\ast}{\pm}}$ mesons at LHCb.
  (2)
  The PLDCV $D_{s}^{\ast}$ decays are favored by the CKM
  element ${\vert}V_{cs}{\vert}$.
  Their branching ratios are about ${\cal O}(10^{-6})$.
  The PLDCV decays $D_{d,s}^{\ast}$ ${\to}$ $e^{-}\bar{\nu}_{e}$,
  ${\mu}^{-}\bar{\nu}_{\mu}$, ${\tau}^{-}\bar{\nu}_{\tau}$
  could be carefully studied at the Belle-II, SCTF or STCF,
  CEPC, FCC-ee, LHCb experiments.
  (3)
  For the $B_{u}^{\ast}$ mesons below the $B{\pi}$ thresholds
  and the $B_{c}^{\ast}$ mesons below both $BD$ and $B_{c}{\pi}$
  thresholds, they decay predominantly through the
  magnetic dipole transitions.
  The branching ratios of the PLDCV $B_{c}^{\ast}$ decays
  favored by the CKM element ${\vert}V_{cb}{\vert}$ could reach
  up to ${\cal O}(10^{-6})$.
  The PLDCV decays $B_{u,c}^{\ast}$ ${\to}$ $e^{-}\bar{\nu}_{e}$,
  ${\mu}^{-}\bar{\nu}_{\mu}$, ${\tau}^{-}\bar{\nu}_{\tau}$
  might be searched for at the CEPC, FCC-ee, LHCb experiments.
  Our rough estimations and findings are summed in Table
  \ref{tab:sum}. We wish that our investigation could provoke
  physicists' researching interest in PLDCV and offer
  a ready reference for the future experimental analysis.

   \begin{table}[ht]
   \caption{The probabilities of experimetal investigation
   on PLDCV, where ${\cal B}$ denotes the branching
   ratio, and the symbol ${\star}$ denotes that the PLDCV
   process might be experimentally accessible in the future.}
   \label{tab:sum}
   \begin{ruledtabular}
   \begin{tabular}{lcccccc}
   decay modes & ${\cal B}$ & Belle-II & SCTF/STCF
   & CEPC & FCC-ee & LHCb \\ \hline
     ${\rho}^{-}$ ${\to}$
     $e^{-}\bar{\nu}_{e}$,
     ${\mu}^{-}\bar{\nu}_{\mu}$
   & ${\cal O}(10^{-13})$
   & & & & & ${\star}$ \\
     $K^{{\ast}-}$ ${\to}$
     $e^{-}\bar{\nu}_{e}$,
     ${\mu}^{-}\bar{\nu}_{\mu}$
   & ${\cal O}(10^{-13})$
   & & & & & ${\star}$ \\
     $D_{d}^{{\ast}-}$ ${\to}$
     $e^{-}\bar{\nu}_{e}$,
     ${\mu}^{-}\bar{\nu}_{\mu}$,
     ${\tau}^{-}\bar{\nu}_{\tau}$
   & ${\cal O}(10^{-10})$
   & ${\star}$ & ${\star}$ & ${\star}$ & ${\star}$ & ${\star}$ \\
     $D_{s}^{{\ast}-}$ ${\to}$
     $e^{-}\bar{\nu}_{e}$,
     ${\mu}^{-}\bar{\nu}_{\mu}$,
     ${\tau}^{-}\bar{\nu}_{\tau}$
   & ${\cal O}(10^{-6})$
   & ${\star}$ & ${\star}$ & ${\star}$ & ${\star}$ & ${\star}$ \\
     $B_{u}^{{\ast}-}$ ${\to}$
     $e^{-}\bar{\nu}_{e}$,
     ${\mu}^{-}\bar{\nu}_{\mu}$,
     ${\tau}^{-}\bar{\nu}_{\tau}$
   & ${\cal O}(10^{-10})$
   & & & ${\star}$ & ${\star}$ & ${\star}$ \\
     $B_{c}^{{\ast}-}$ ${\to}$
     $e^{-}\bar{\nu}_{e}$,
     ${\mu}^{-}\bar{\nu}_{\mu}$,
     ${\tau}^{-}\bar{\nu}_{\tau}$
   & ${\cal O}(10^{-6})$
   & & & ${\star}$ & ${\star}$ & ${\star}$
  \end{tabular}
  \end{ruledtabular}
  \end{table}

  \section*{Acknowledgments}
  The work is supported by the National Natural Science Foundation
  of China (Grant Nos. 11705047, 11981240403, U1632109, 11547014),
  the Chinese Academy of Sciences Large-Scale Scientific Facility
  Program (1G2017IHEPKFYJ01) and the Program for Innovative Research
  Team in University of Henan Province (19IRTSTHN018).
  We thank Prof. Haibo Li (IHEP@CAS),
  Prof. Shuangshi Fang (IHEP@CAS),
  Prof. Frank Porter (Caltech),
  Prof. Antimo Palano (INFN),
  Prof. Chengping Shen (Fudan University),
  Dr. Xiao Han (Fudan University),
  Prof. Xiaolin Kang (China University of Geosciences),
  Ms. Qingping Ji (Henan Normal University),
  Ms. Huijing Li (Henan Normal University)
  for their kindly help and valuable discussion.


  \end{document}